\documentclass[conference]{IEEEtran}
\IEEEoverridecommandlockouts
\usepackage{booktabs}
\usepackage{helvet}  
\usepackage{courier}  
\usepackage{url}  
\usepackage{graphicx}  
\usepackage{algorithm}
\usepackage{algpseudocode}
\usepackage{subcaption}
\usepackage{color}
\usepackage{amsmath} 
\usepackage{amssymb}  
\usepackage{mathtools}
\usepackage{cleveref}
\usepackage{xcolor}
\usepackage{cleveref}
\usepackage{booktabs}
\usepackage{adjustbox}

\def\BibTeX{{\rm B\kern-.05em{\sc i\kern-.025em b}\kern-.08em
    T\kern-.1667em\lower.7ex\hbox{E}\kern-.125emX}}

\usepackage{etoolbox}
\newbool{widetables}

\begin{document}
\title{Multi-Agent Reinforcement Learning for Problems with Combined Individual and Team Reward \\
\thanks{This work had been supported in part
by the National Science Foundation under grant number IIS-1409823.}
}

\author{\IEEEauthorblockN{Hassam Ullah Sheikh}
\IEEEauthorblockA{\textit{Department of Computer Science} \\
\textit{University of Central Florida}\\
Orlando, USA \\
hassam.sheikh@knights.ucf.edu}
\and
\IEEEauthorblockN{Ladislau B{\"o}l{\"o}ni}
\IEEEauthorblockA{\textit{Department of Computer Science} \\
\textit{University of Central Florida}\\
Orlando, USA \\
lboloni@cs.ucf.edu}
}
\maketitle

\begin{abstract}
Many cooperative multi-agent problems require agents to learn individual tasks while contributing to the collective success of the group. This is a challenging task for current state-of-the-art multi-agent reinforcement algorithms that are designed to either maximize the global reward of the team or the individual local rewards. The problem is exacerbated when either of the rewards is sparse leading to unstable learning. To address this problem, we present \textit{Decomposed Multi-Agent Deep Deterministic Policy Gradient (DE-MADDPG)}: a novel cooperative multi-agent reinforcement learning framework that simultaneously learns to maximize the global and local rewards. We evaluate our solution on the challenging defensive escort team problem and show that our solution achieves a significantly better and more stable performance than the direct adaptation of the MADDPG algorithm.
\end{abstract}


\section{Introduction}
\label{sec:Introduction}

Cooperative multi-agent problems are prevalent in real-world settings such as strategic conflict resolution~\cite{Leibo-2017-AAMAS}, coordination between autonomous vehicles~\cite{CAO-2012-TII} and collaboration of agents in defensive escort teams~\cite{Sheikh-2019-COMPSAC}. Such problems can be modelled as dual-interest: each agent is simultaneously working towards maximizing its own payoff (local reward) as well as the collective success of the team (global reward). For example, autonomous vehicles in double-lane merge conflicts must perform cooperative maneuvers without diverging from their destination-bound nominal trajectories. Similarly, in the case of a defensive escort team, each agent has to maintain a specific distance from the payload to avoid disrupting any social norms without sacrificing the security of the payload. Despite the recent success of multi-agent reinforcement learning (MARL) in multiplayer games like Dota 2~\cite{OpenAI-2018}, Quake III Capture-the-Flag~\cite{Max-2018-ARXIV} and Starcraft II~\cite{Vinyals-2019-Nature} or learning to use tools~\cite{Baker-2019-ARXIV}, learning multi-agent cooperation while simultaneously maximizing local rewards is still an open challenge. In this learning problem, to which we will refer as ``\textbf{dual-reward MARL}", the agents are {\em explicitly} receiving two reward signals: the global team reward and the agent's individual local reward. 

Current state-of-the-art MARL algorithms can be categorized in two types. For algorithms such as COMA~\cite{Foerster-2018-AAAI} and QMIX~\cite{Rashid-2018-ICML}, the goal is to maximize the global reward for the success of the group while algorithms such MADDPG~\cite{Lowe-2017-NIPS} and M3DDPG~\cite{Li-2019-AAAI} focus on optimizing local rewards without any explicit notion of coordination. As shown in~\cite{Yang-2018-CORR} and in our findings in \Cref{sec:experiments}, a direct adaptation of these algorithms to dual-reward problems often leads to poor performance and unstable learning. Generally, these adaptations happen in the reward function space where the local and the global reward signals are combined to form an entangled multi-objective reward function~\cite{Sheikh-2018-GoalsRL}. This coupling of reward functions leads to two problems. First, the entangled reward function becomes unfactorizable during training, causing the learning to oscillate between optimizing either the global or the local reward leading to a sub-optimal and unstable solution. This problem is exacerbated when either of the rewards is sparse, thus, leading to a bias towards the other. The second problem is that maximizing the entangled reward function does not correspond to maximizing the objective function.

To address these issues, we present \textit{Decomposed Multi-Agent Deep Deterministic Policy Gradient (DE-MADDPG)}: a novel cooperative multi-agent reinforcement learning framework built on top of deterministic policy gradients that simultaneously learns to maximize the global and the local rewards without the need of creating an entangled multi-objective reward function. The core idea behind DE-MADDPG is to train two critics. The \textit{global critic}, shared between all the cooperating agents takes as input the observations and actions of these agents and estimates the sum of the global expected reward. The \textit{local critic} receives as input only the observation and action of the particular agent and estimates the sum of local expected reward. The advantage of training two critics is that the step of designing an entangled multi-objective reward function can be skipped altogether.

\begin{figure*}[t]
    \centering
    \begin{subfigure}[t]{0.24\textwidth}
        \centering
        \frame{\includegraphics[height=1.55in]{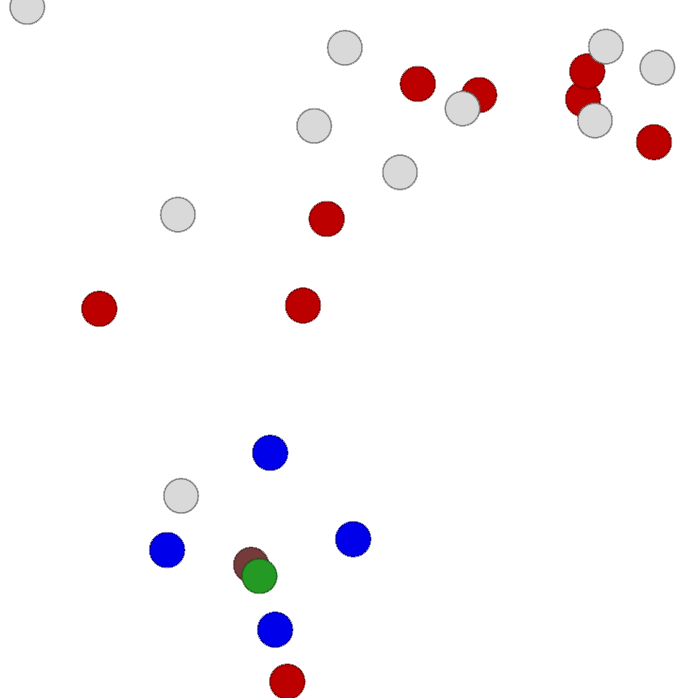}}
        \caption{Random Landmarks}
    \end{subfigure}%
    ~
    \begin{subfigure}[t]{0.24\textwidth}
        \centering
        \frame{\includegraphics[height=1.55in]{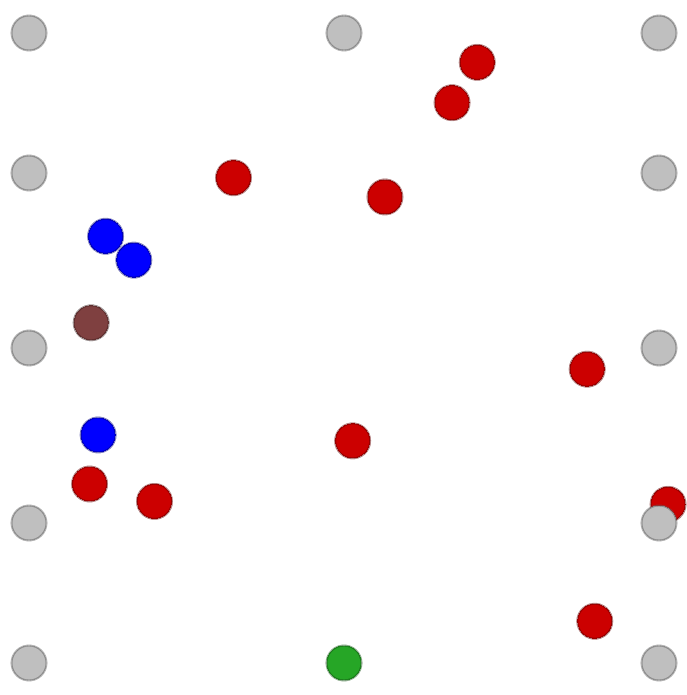}}
        \caption{Shopping Mall}
    \end{subfigure}%
    ~
    \begin{subfigure}[t]{0.24\textwidth}
        \centering
        \frame{\includegraphics[height=1.55in]{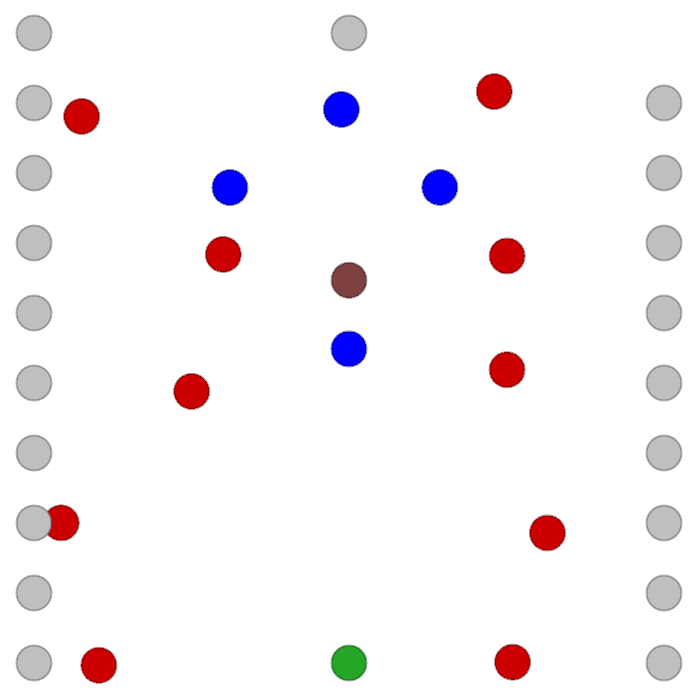}}
        \caption{Street}
    \end{subfigure}%
    ~
    \begin{subfigure}[t]{0.24\textwidth}
        \centering
        \frame{\includegraphics[height=1.55in]{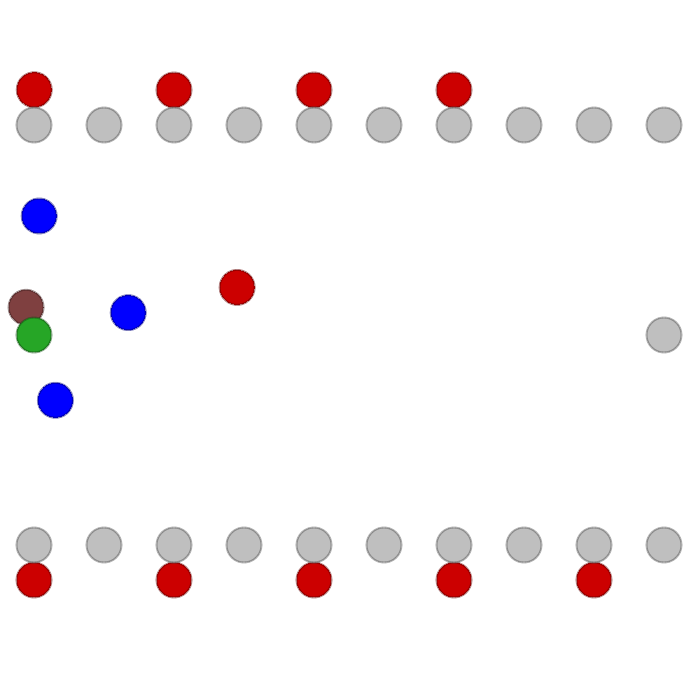}}
        \caption{Pie-in-the-Face}
    \end{subfigure}
    \caption{Four environments for the defensive escort team problem~\cite{Sheikh-2019-AAMAS}. The team of bodyguards (blue) need to protect the VIP (brown) in the environments, from left to right: "Random Landmarks", "Shopping Mall", "Street" and "Pie-in-the-Face".}
    \label{fig:environments}
\end{figure*}

To summarize, our contributions in this paper are the following:

\begin{itemize}
    \item We develop a dual-critic framework for multi-agent reinforcement learning that learns to simultaneously maximize the decomposed global and local rewards.
    \item Taking advantage of the decomposition, we treat the global critic as a single-agent critic. This allows us to apply performance enhancement techniques such as Prioritized Experience Replay (PER)~\cite{Schaul-2016-ICLR} and Twin Delayed Deep Deterministic Policy Gradients (TD3)~\cite{Fujimoto-2018-ICML} to tackle the overestimation bias problem in Q-functions. This was not previously feasible in the multi-agent RL setting.
    \item We evaluate our proposed solution on the defensive escort team problem~\cite{Sheikh-2019-COMPSAC,Sheikh-2019-AAMAS} (see Figure~\ref{fig:environments}) and show that it achieves a significantly better and more stable performance than the direct adaptation of the MADDPG algorithm.
\end{itemize}

\section{Related work}

Early theoretical work in MARL was limited to discrete state and action spaces 
\cite{Tan-1993-ICML,Littman-1994-Markov,Hu-2003-JMLR}.
Recent work have adopted techniques from single-agent deep RL to develop general algorithms for high-dimensional continuous space environments requiring complex agent interactions \cite{Leibo-2017-AAMAS, Mordatch-2017-AAAI,Lowe-2017-NIPS}.

Cooperative multi-agent learning is important since many real-world problems can be formulated as distributed systems with decentralized agents that must coordinate to achieve shared objectives~\cite{Panait-2005-AAMAS}. Similar to our work, \cite{Austerweil-2015-AAAI} have shown that agents whose rewards depend on all agents' success perform better than agents who optimize for their own success.
In the special case when all agents have a single goal and share a global reward, COMA~\cite{Foerster-2018-AAAI} uses a counterfactual baseline.
However, the centralized critic in these methods only focuses on optimizing the collective success of the group. 
When a global objective is the sum of agents' individual objectives,
value-decomposition methods optimize a centralized Q-function while preserving scalable decentralized execution~\cite{Rashid-2018-ICML}, but do not address credit assignment.
While MADDPG~\cite{Lowe-2017-NIPS} and M3DDPG~\cite{Li-2019-AAAI} apply to agents with different reward functions, they do not specifically address the need for cooperation; in fact, they do not distinguish the problems of cooperation and competition, despite the fundamental difference.

To the best of our knowledge, dual-reward MARL was not explicitly addressed in the existing literature. Among the related problems, ~\cite{Zhang-2018-ICML} explored the multi-goal problem and analyzed its convergence in a special networked setting restricted to fully-decentralized training. In contrast, we are conducting centralized training with decentralized execution.
In contrast to multi-\textit{task} MARL, which aims for generalization among \textit{non-simultaneous} tasks~\cite{Omidshafiei-2017-ICML},
and in contrast to hierarchical methods with top-level managers that \textit{sequentially} select subtasks~\cite{Vezhnevets-2017-ICML}, our decentralized agents must cooperate \textit{in parallel} to successfully achieve the global and their respective local objectives.

\section{Background}
\label{sec:Background}
\subsection{Policy Gradients}

Policy gradient methods have been shown to learn the optimal policy in a variety of reinforcement learning tasks. The main idea behind policy gradient methods is that instead of parameterizing the Q-function to extract the policy, we parameterize the policy using the parameters $\theta$ to maximize the objective represented as $J\left(\theta\right)=\mathbb{E}\left[\mathbb{R}^{t}\right]$ by taking a step in the direction
\[
\nabla J{\left(\theta\right)}=\mathbb{E}\left[\nabla_{\theta}\log\pi_{\theta}\left(a|s\right)Q^{\pi}\left(s,a\right)\right]
\]
Policy gradient methods are prone to the high variance problem. Several methods such as~\cite{Wu-2018-ICLR,Schulman-2017-CORR} have been shown to reduce the variability by introducing a {\em critic}, a Q-function that tells about the goodness of a reward by working as a baseline. \cite{Silver-2014-ICML} has shown that it is possible to extend the policy gradient framework to deterministic policies {\em i.e.} $\pi_{\theta}:\mathcal{S}\rightarrow\mathcal{A}$.
In particular we can write $\nabla J\left(\theta\right)$ as
\[
\nabla J{\left(\theta\right)}=\mathbb{E}\left[\nabla_{\theta}\pi\left(a|s\right)\nabla_{a}Q^{\pi}\left(s,a\right)|_{a=\pi\left(s\right)}\right]
\]
A variation of this model, Deep Deterministic Policy Gradients (DDPG) ~\cite{Lillicrap-2015-ICLR} is an off-policy algorithm that approximates the policy $\pi$ and the critic $Q^{\pi}$  with deep neural networks. DDPG uses an experience replay buffer alongside a target network to stabilize the training. Twin Delayed Deep Deterministic Policy Gradients (TD3)~\cite{Fujimoto-2018-ICML} improves on DDPG by addressing the overestimation bias of the Q-function, similarly to Double Q-learning~\cite{Van-2016-AAAI}. They find that approximation errors of the neural network, combined with gradient descent make DDPG tend to overestimate the Q-values, leading to a slower convergence. TD3 addresses this by using two Q-networks $Q_{\psi_1},Q_{\psi_2}$, along with two target networks. The Q-functions are updated with the target $y=r^t + \gamma \min_{1,2} Q_{\psi_i'}(s'^t,a'^t)$, while updating the policy with $Q_{\psi_1}$.
Additionally, they introduce target policy smoothing by adding noise in the determination of the next action for the critic target $a'^t = \mu_{\theta_\pi'}(s') + \epsilon$, with $\epsilon$ being clipped Gaussian noise $\epsilon = \mathtt{clip} (\mathcal{N}(0,\sigma),-c,c)$, where $c$ is a tunable parameter. Additionally, they use delayed policy updates and only update the policy $\pi$ and target network parameters once every $d$ critic updates.

Multi-agent deep deterministic policy gradients (MADDPG)~\cite{Lowe-2017-NIPS} extends DDPG for the multi-agent setting where each agent has it's own policy. The gradient of each policy is written as
\[
\nabla J{\left(\theta_{i}\right)} = \mathbb{E}\left[\nabla_{\theta_{i}}\pi_{i}\left(a_{i}|o_{i}\right)\nabla _{a_i}Q_{i}^{\pi}\left(s,a_{1},\ldots,a_{N}\right)|_{a_i=\pi_i\left(o_i\right)}\right]
\]
\noindent where $s=\left(o_1, \ldots, o_N\right)$  and   $Q_{i}^{\pi}\left(s,a_{1},\ldots,a_{N}\right)$ is a centralized action-value function that takes the actions of all the agents in
addition to the state of the environment to estimate the Q-value for agent
$i$. Since every agent has its own Q-function, the model allows the agents
to have different action space and reward functions. The primary insight
behind MADDPG is that knowing all the actions of other agents makes
the environment stationary, even though their policy changes.

Very recently~\cite{Ackermann-2019-NIPS} proposed Multi-Agent TD3 (MATD3) extending MADDPG by replacing the deterministic policy gradients with twin delayed deterministic policy gradient to tackle the overestimation bias problem.

\begin{figure}
    \centering
    \includegraphics[height=1.5in, width=2.5in]{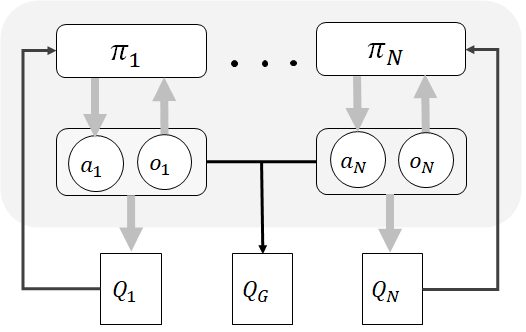}
    \caption{An overview of the Decomposed Multi-Agent Deep Deterministic Policy Gradient architecture. In contrast to MADDPG which uses a  single centralized critic, DE-MADDPG has a global centralized critic shared between all cooperating agents and a local critic specific to the agent.}
\label{fig:demaddpg}
\end{figure}

\section{Decomposed Multi-Agent Deep Deterministic Policy Gradient}
We propose {\em Decomposed Multi-Agent Deep Deterministic Policy Gradient}: a multi-agent deep reinforcement learning algorithm that learns to simultaneously maximize the group's global reward and the agent's local rewards. Our approach uses a two critic approach to train policies and value functions that are optimal to maximize the global and local rewards respectively.

The main idea is to combine MADDPG (or MATD3) for maximizing global rewards with a standard single agent DDPG (or TD3). Intuitively, the goal is to move the policy in the direction that maximizes both the global and the local critic. The resulting learning paradigm is similar to the centralized training with decentralized execution during testing used by~\cite{Lowe-2017-NIPS}. In this setting, additional information is provided for the agents during training that is not available during test time. 


Concretely, we consider an environment with $N$ agents with policies ${\pi}=\{{\pi_1}, \ldots, {\pi_N}\}$ parameterized by ${\theta}=\{{\theta_1}, \ldots, {\theta_N}\}$. The \textit{multi-agent deep deterministic policy gradient} for agent $i$ can written as
\[
\nabla J{\left(\theta_{i}\right)}=\mathbb{E}\left[\nabla_{\theta_{i}}\pi_{i}\left(a_{i}|o_{i}\right)\nabla _{a_i}Q_{i}^{\pi}\left(s,a_{1},\ldots,a_{N}\right)|_{a_i=\pi_i\left(o_i\right)}\right]
\]
\noindent where $s=\left(o_1, \ldots, o_N\right)$  and   $Q_{i}^{\pi}\left(s,a_{1},\ldots,a_{N}\right)$ is a centralized action-value function parameterized by $\phi_i$ that takes the actions of all the agents in addition to the state of the environment to estimate the Q-value for agent $i$. We extend the idea of MADDPG by introducing a local critic. Now the modified policy gradient for each agent $i$ can be written as
\begin{align}
\label{eq:demaddpg}
\begin{split}
 \nabla J{\left(\theta_{i}\right)} &= \overbrace{\mathbb{E}_{s,a\thicksim \mathcal{D}}\Bigg[\nabla_{\theta_{i}}\pi_{i}\left(a_i|o_{i} \right) \nabla_{a_i}Q_{\psi}^{g}\left(s,a_{1},\ldots,a_{N}\right)  \Bigg]}^{MADDPG}\\
 &+  \mathbb{E}_{o_i,a_i \thicksim \mathcal{D}}\Bigg[\nabla_{\theta_{i}}\pi_{i}\left(a_i|o_{i} \right) \nabla_{a_i}Q_{i}^{\pi}\left(o_i,a_{i}\right)  \Bigg]\Bigg\}DDPG
\end{split}
\end{align}
\noindent where ${a_i=\pi_i\left(o_i\right)}$ is action from agent $i$ following policy $\pi_i$ and $\mathcal{D}$ is the experience replay buffer. The global critic is $Q_{\psi}^{g}$ is updated as:
\begin{equation*}
\mathcal{L}\left(\psi\right)=\mathbb{E}_{s,a,r, s'}\left[\left(  Q_{\psi}^{g}\left(s,a_{1},\ldots,a_{N}\right)-y_g\right)^2  \right]
\end{equation*}
\noindent where $y_g$ is defined as:
\begin{equation*}
y_g=r_g+\gamma Q_{\psi'}^{g}\left(s',a'_{1},\ldots,a'_{N}\right)|_{a'_i=\pi'_i\left(o_i'\right)}
\end{equation*}
\noindent where ${\pi'}=\{{\pi'_1}, \ldots, {\pi'_N}\}$ are target policies parameterized by ${\theta'}=\{{\theta_1'}, \ldots, {\theta'_N}\}$. Similarly, The local critic is $Q_{i}^{\pi}$ is updated as:
\begin{equation*}
\mathcal{L}\left(\phi_i\right)=\mathbb{E}_{o,a,r, o'}\left[\left(  Q_{i}^{\pi}\left(o_i, a_{i}\right)- y_l\right)^2  \right]
\end{equation*}
\noindent where $y_l$ is defined as:
\begin{equation*}
y_l=r^i_l+\gamma Q_{\phi'_i}^{\pi'}\left(o_i',a'_{i}\right)|_{a'_i=\pi'_i\left(o_i'\right)}
\end{equation*}
Overestimation bias in Q-functions have been thoroughly studied in~\cite{Fujimoto-2018-ICML,Ackermann-2019-NIPS}. This overestimation bias can be problemsome in multi-agent settings especially in real time autonomous systems. For example, in resolving the double lane merge conflict in autonomous vehicles, the vehicles might consider the current state to be near conflict resolution thus taking a dangerous turns. To solve this problem~\cite{Fujimoto-2018-ICML} have proposed a double critic approach to minimize the overestimation bias. Motivated from the results in~\cite{Fujimoto-2018-ICML}, we replace the Multi-Agent Deterministic Policy Gradient of the global critic in~\Cref{eq:demaddpg} with Twin Delayed Deterministic Policy Gradient. Therefore our updated policy gradient becomes

\begin{align}
\label{eq:demaddpg_td3}
\begin{split}
 \nabla J{\left(\theta_{i}\right)} &= \mathbb{E}_{s,a\thicksim \mathcal{D}}\Bigg[\nabla_{\theta_{i}}\pi_{i}\left(a_i|o_{i} \right) \nabla_{a_i}Q_{\psi_1}^{g_1}\left(s,a_{1},\ldots,a_{N}\right)  \Bigg]\\
 &+  \mathbb{E}_{o_i,a_i \thicksim \mathcal{D}}\Bigg[\nabla_{\theta_{i}}\pi_{i}\left(a_i|o_{i} \right) \nabla_{a_i}Q_{i}^{\pi}\left(o_i,a_{i}\right)  \Bigg]
\end{split}
\end{align}

The twin global critics are updated as
\begin{equation*}
\mathcal{L}\left(\psi_i\right)=\mathbb{E}_{s,a,r, s'}\left[\left(  Q_{\psi_i}^{g_i}\left(s,a_{1},\ldots,a_{N}\right)-y_g\right)^2  \right]
\end{equation*}

\noindent where $y_g$ is defined as:
\begin{equation*}
y_g=r_g+\gamma \min_{i=1,2}Q_{\psi_i'}^{g_i}\left(s',a'_{1},\ldots,a'_{N}\right)|_{a'_i=\pi'_i\left(o_i'\right)}
\end{equation*}
\newcommand{\Mod}[1]{\ \mathrm{mod}\ #1}
\begin{algorithm}
  \caption{Decomposed Multi-agent Deep Policy Gradient}
  \begin{algorithmic}[1]
  \State Initialize main global critic networks $Q_{\psi}^{g_1}$ and $Q_{\psi}^{g_2}$.
  \State Initialize target global critic networks $Q_{\psi'}^{g_1}$ and $Q_{\psi'}^{g_2}$.
  \State Initialize each agents policy and critic networks.
  \For{episode = 1 to $T$}
    \For{t = 1 to episode--length}
        \State Get environment state $s^t$.
   		\State For each agent $i$, select action $a_{i}^{t}=\pi_{\theta_{i}}\left(o^{t}_{i}\right)$
        \State Execute actions $\textbf{a}^{t}=\left[a^{t}_{1},\ldots,a^{t}_{N}\right]$
        \State Receive global $r_{g}^{t}$ and local rewards $\textbf{r}^{t}_{l}$.
        \State Store $\left( {s}^{t}, \textbf{a}^{t}, \textbf{r}^{t}_{l}, {r}^{t}_{g}, {s}^{t+1}\right)$ in replay buffer.
    \EndFor
    \State /* Train global critic*/
    \State Sample minibatch of size S $\left(\textbf{s}^{j}, \textbf{a}^{j}, \textbf{r}_{g}^{j}, \textbf{s}^{'j}\right)$ from buffer. 
    \State $\left(a'_{1},\ldots,a'_{N} \right) \coloneqq \left(\pi'_{\theta_{i}}(o'^{j}_{i}),\ldots,\pi'_{\theta_{N}}(o'^{j}_{N}) \right)$
    \State Set $y^{j}_{g} = {r}_{g}^{j}+\gamma \min_{i=1,2} Q_{\psi'_i}^{g_i}\left({s}^{'j},a'_{1},\ldots,a'_{N}\right)$
    \State Update global critics by minimizing  \[ \frac{1}{S} \sum_j \left(y^{j}_{g} -  Q_{\psi_i}^{g_i}\left({s}^{j},a^{j}_{1},\ldots,a^j_{N}\right) \right)^2\]
    \State Update target network parameters
        \begin{align*}
        	\psi'_i\leftarrow \tau\psi_i + \left(1-\tau\right)\psi'_i
        \end{align*}
    \If{$\text{episode}\Mod{d}$}
    \State /* Train local critics and update agent policies*/
    \For{agent $i = 1$ to $N$}
    \State Sample minibatch of size S $\left(\textbf{s}^{j}, \textbf{a}^{j}, \textbf{r}^{j}_l, \textbf{s}^{'j}\right)$
    \State Set $y^j = r_{i_{l}}^{j}+\gamma Q_{\phi'_i}^{\pi'}\left({o}^{'j},\pi'_{\theta_{i}}({o}^{'j}_{i}) \right)$
        \State Update local critic by minimizing  \[ \frac{1}{S} \sum_j \left(y^j -  Q_{\phi_i}^{\pi}\left({o}^{j},a^{j}_{i} \right) \right)^2\]
         \State 
         \begin{align*}
         \theta_i = \theta_i + \frac{1}{S} \displaystyle \sum_j  \nabla_{\theta_{i}}\pi_{i}\left(a_{i}|o_{i}^j\right)\nabla _{a_i} Q_{\psi}^{g_1}\left({s}^{j},a^{j}_{1},\ldots,a^j_{N}\right) \\+ \nabla_{\theta_{i}}\pi_{i}\left(a_{i}|o_{i}^j\right)\nabla _{a_i} Q_{\phi_i}^{\pi}\left({o}^{j},a^{j}_{i}\right)
         \end{align*}{}
        \EndFor
        \State Update target network parameters for each agent $i$
        \begin{align*}
        	\theta'_i \leftarrow \tau\theta_i + \left(1-\tau\right)\theta'_i\\
        \phi'_i \leftarrow \tau\phi_i + \left(1-\tau\right)\phi'_i
        \end{align*}
    \EndIf
    \EndFor
  \end{algorithmic}
  \label{alg:1}
\end{algorithm}
Similarly, the local critics can be updated using TD3 update style but for simplicity, we will use the standard DDPG to update the local critics.   
The overall algorithm to which we refer as \textit{Decomposed Multi-Agent Deterministic Policy Gradient (DE-MADDPG)} is described in~\Cref{alg:1}. The overview of the architecture can be seen in~\Cref{fig:demaddpg}.

\begin{figure}
    \centering
    \begin{subfigure}[t]{0.25\textwidth}
        \centering
        \includegraphics[width=1.\linewidth, height=1.60in]{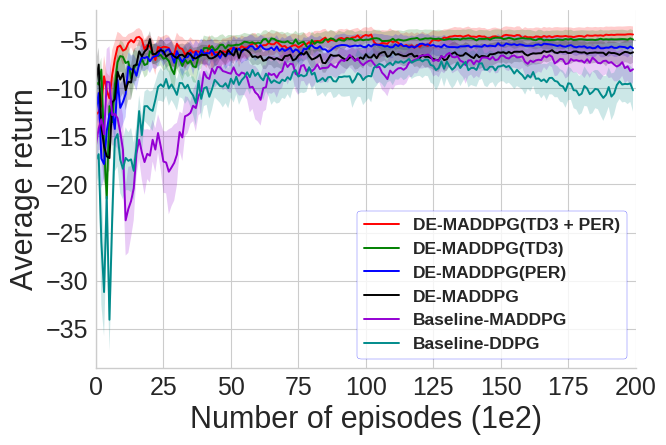}
        \caption{Random Landmarks}

        \label{fig:random_landmark_reward}
    \end{subfigure}%
    ~
    \begin{subfigure}[t]{0.25\textwidth}
        \centering
        \includegraphics[width=1.\linewidth, height=1.60in]{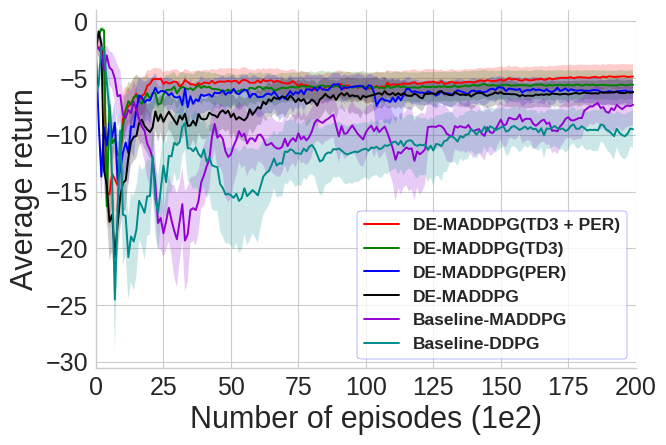}
        \caption{Shopping Mall}
        \label{fig:shopping_mall_reward}
    \end{subfigure}%
    \newline
    \begin{subfigure}[t]{0.25\textwidth}
        \centering
        \includegraphics[width=1.\linewidth, height=1.60in]{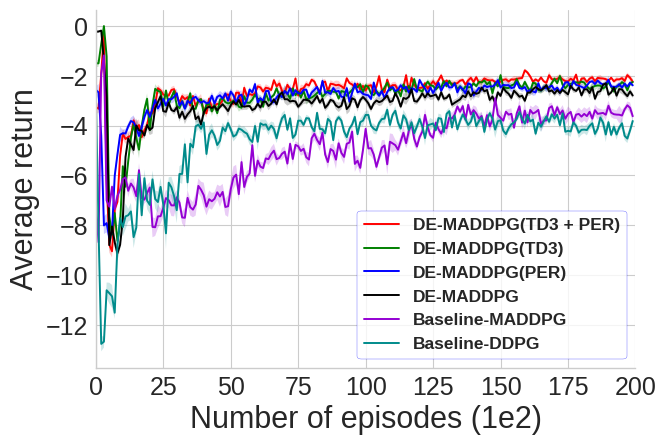}
        \caption{Street}
        \label{fig:street_reward}
    \end{subfigure}%
    ~
    \begin{subfigure}[t]{0.25\textwidth}
        \centering
        \includegraphics[width=1.\linewidth, height=1.60in]{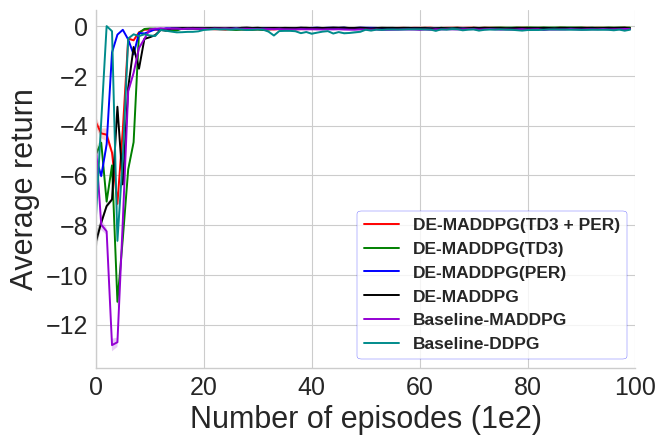}
        \caption{Pie-in-the-Face}
        \label{fig:red_carpet_reward}
    \end{subfigure}
    \caption{Learning curves representing the average cumulative global reward. The higher reward represents higher protection to the VIP (payload).}
    \label{fig:main_results}
\end{figure}

\begin{figure}[t]
    \centering
    \begin{subfigure}[t]{0.25\textwidth}
        \centering
        \includegraphics[width=1.\linewidth, height=1.60in]{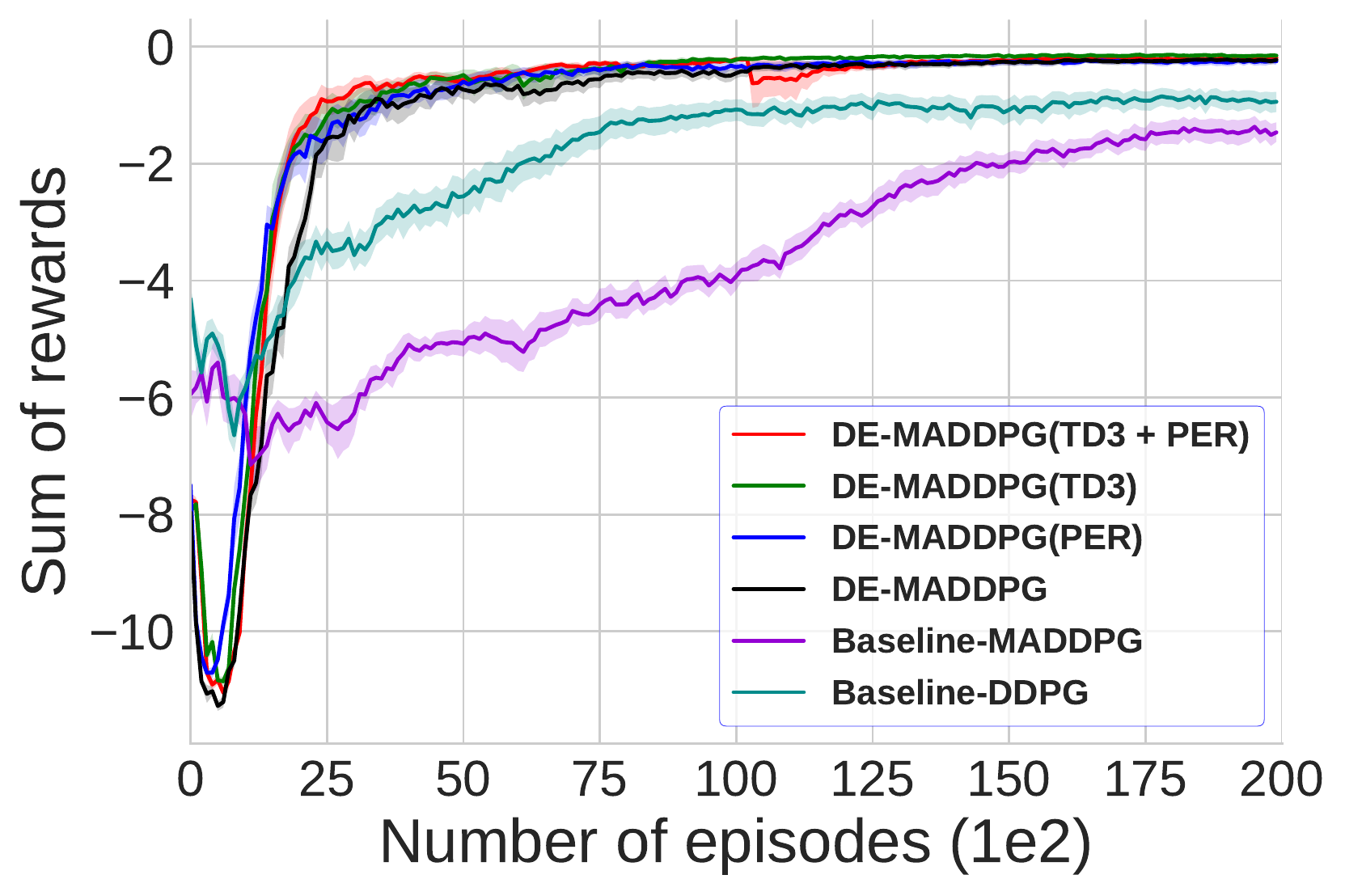}
        \caption{Random Landmarks}

        \label{fig:random_landmark_reward_local}
    \end{subfigure}%
    ~
    \begin{subfigure}[t]{0.25\textwidth}
        \centering
        \includegraphics[width=1.\linewidth, height=1.60in]{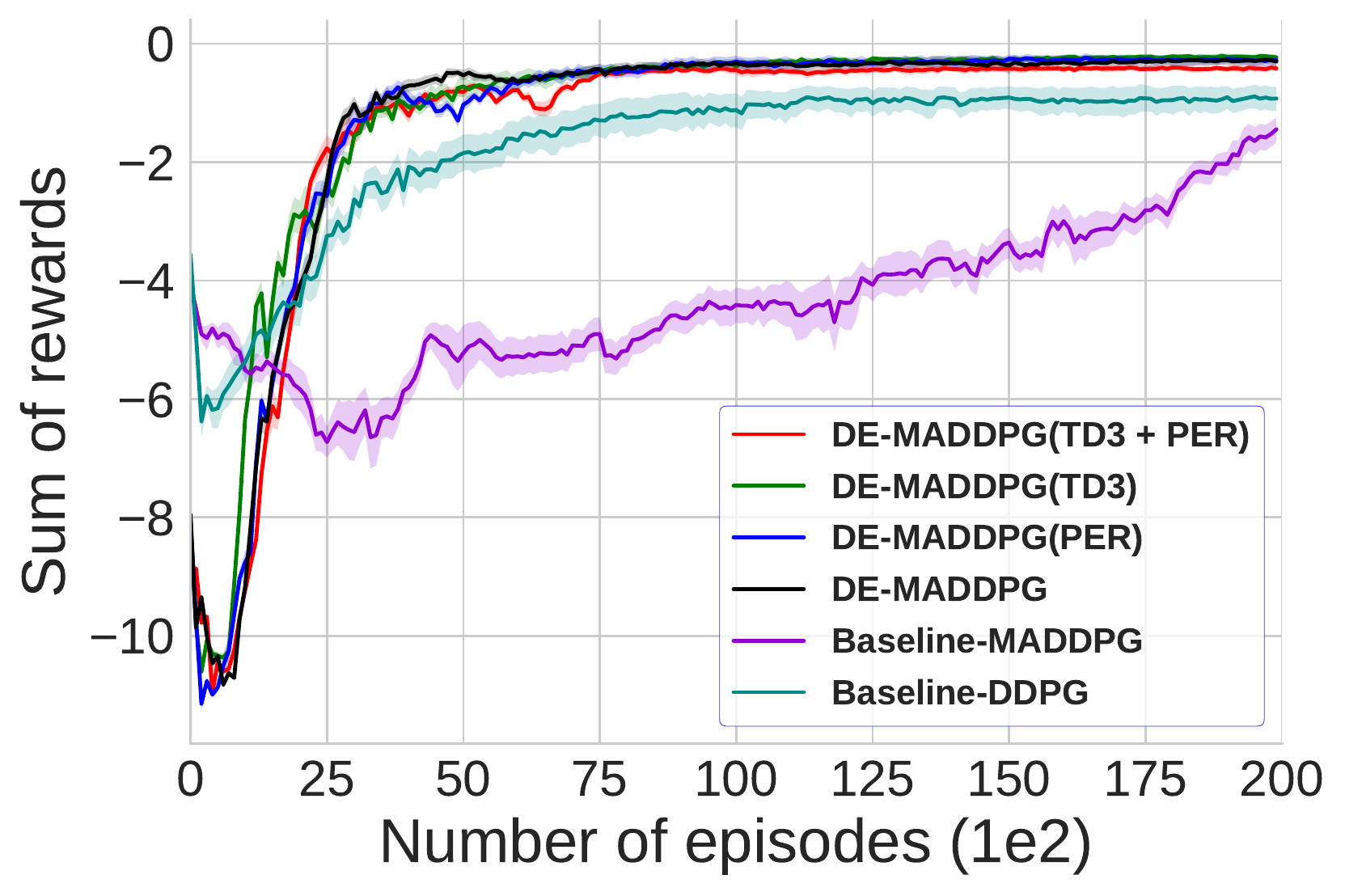}
        \caption{Shopping Mall}
        \label{fig:shopping_mall_reward_local}
    \end{subfigure}%
    \newline
    \begin{subfigure}[t]{0.25\textwidth}
        \centering
        \includegraphics[width=1.\linewidth, height=1.60in]{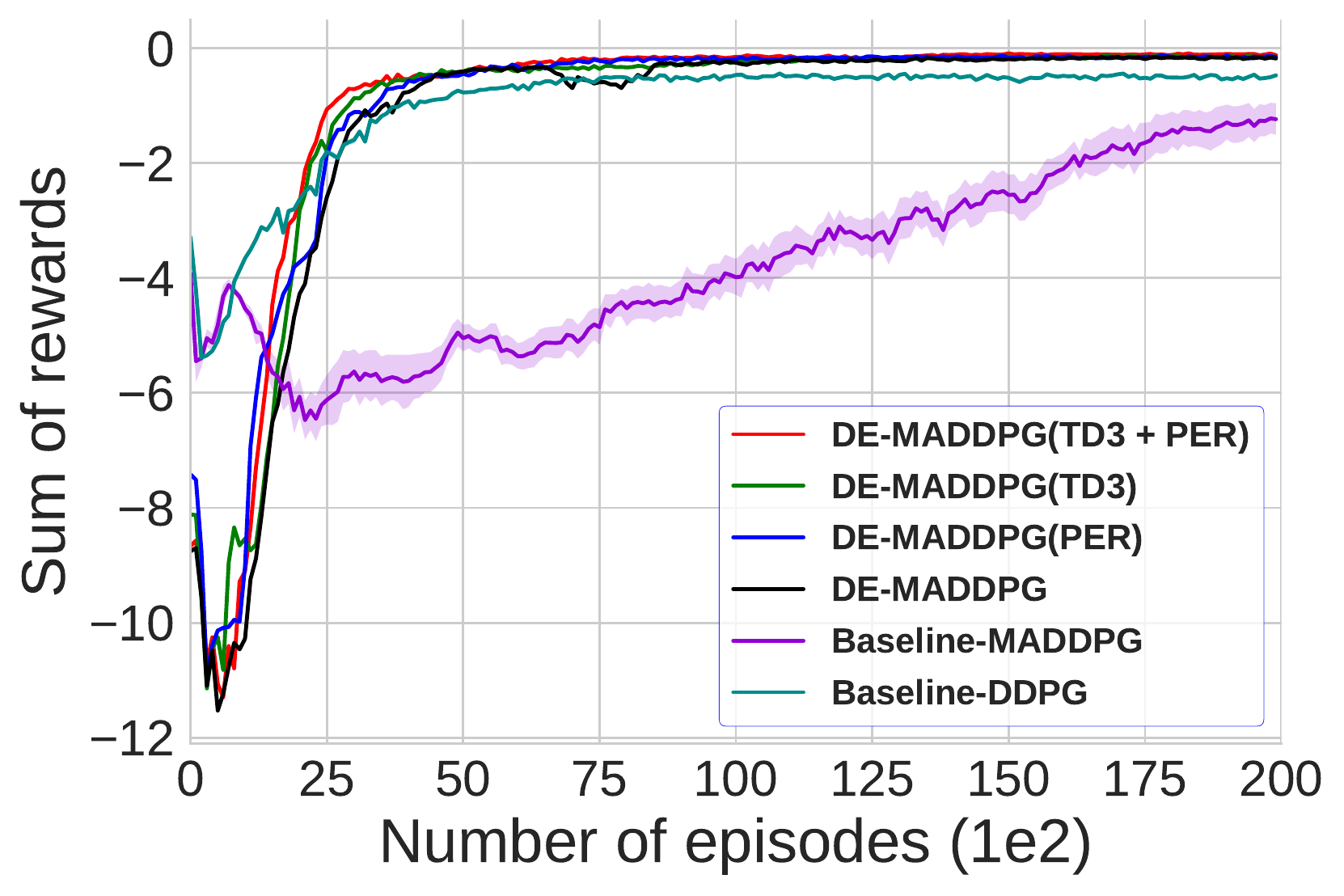}
        \caption{Street}
        \label{fig:street_reward_local}
    \end{subfigure}%
    ~
    \begin{subfigure}[t]{0.25\textwidth}
        \centering
        \includegraphics[width=1.\linewidth, height=1.60in]{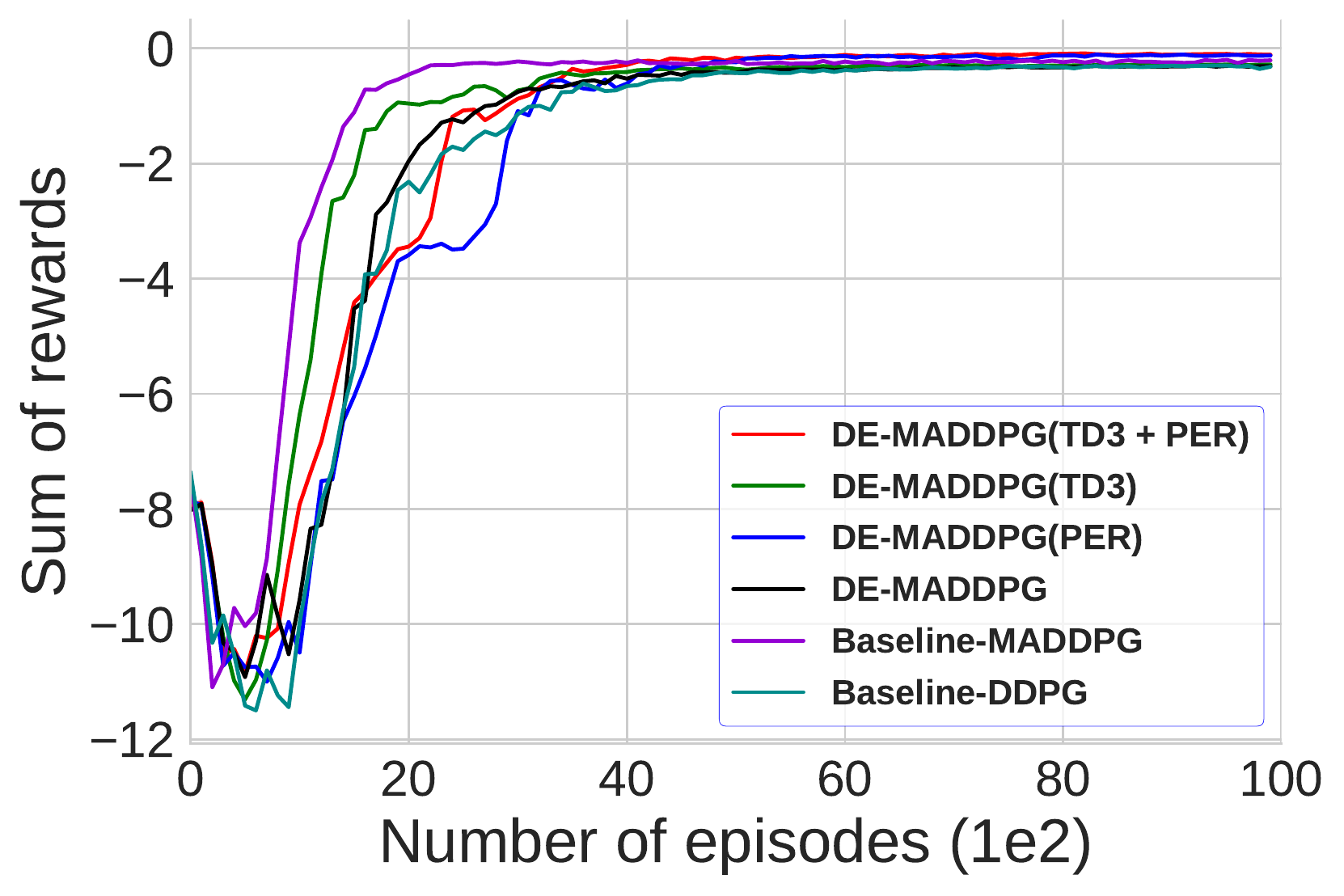}
        \caption{Pie-in-the-Face}
        \label{fig:red_carpet_reward_local}
    \end{subfigure}
    \caption{Learning curves of the experiments representing the sum of local rewards. Notice that similar to global rewards, the local reward learning of MADDPG and DDPG is also unstable and reaches only at sub-optimal performance.}
    \label{fig:main_results_local}
\end{figure}

\section{Experiments}
\label{sec:experiments}



\subsection{Environments}
\label{subsec:env}

We perform our experiments using the defensive escort problem on four VIP protection environments~\cite{Sheikh-2019-AAMAS, Sheikh-2019-COMPSAC}. This is a medium-size collaborative problem where a defensive escort team of agents is learning to maintain an optimal formation around the VIP (payload). The objective of the defenders is to minimize the potential physical attacks as the VIP is moving in a variety of different real world scenarios and are implemented in the Multi-Agent Particle Environment~\cite{Mordatch-2017-AAAI}. An illustration of the environments can be seen in~\Cref{fig:environments}.

The environment consists of the VIP (payload), defensive escort team and one or more classes of bystanders. The VIP (brown disk) starts from the starting point and moves towards the destination landmark (green disk). The goal of the defensive escort team is learn an optimal formation around the VIP to protect it from potential physical attacks. In order to maintain social norms, the defensive agents need to maintain a certain distance from the VIP. To closely simulate real world situations and providing substantial variability, four different scenarios were developed that differ in the number, arrangement of the landmarks and the behavior of the different classes of bystanders:

\begin{itemize}
\item \textbf{Random Landmark:}  In this fully stochastic scenario, landmarks are placed randomly in the area. The starting point and destination for the VIP are randomly selected landmarks. The bystanders are performing random waypoint navigation: they choose a landmark at random, move towards it and repeat this process till the end of the episode.

\item \textbf{Shopping Mall:} In this scenario, landmarks are placed at the edge of the environment emulating shops on a street or shopping mall. The bystanders are moving between randomly chosen shops.

\item \textbf{Street:} This scenario aims to model a crowded sidewalk. The bystanders are moving in two different directions towards waypoints that are outside the current area. However, due to their proximity to each other, the position of the other bystanders influence their movement described by laws of particle motion~\cite{Vicsek-1995-PRL}.


\item \textbf{Pie-in-the-Face}: This scenario models a VIP walking a ``red carpet'', with bystanders standing behind designed lines. In this scenario, an  unruly bystander breaks the line to approach the VIP (presumably, to throw a pie in his/her face).
\end{itemize}

The state of the environment is given by the locations of the landmarks, bystanders, VIP and the defensive team. To closely represent a real world bodyguard that has a limited range of perception, the observation of each agent is the relative physical state of the nearest $M$ bystanders, the VIP and the remaining members of the defensive escort team and represented as $o_{i}=\left[x_{j,\ldots N+M} \right]\in \mathcal{O}_i$ where $x_{j}$ is the observation of the entity $j$ from the perspective of agent $i$. In our experiments, we used $M=5$.

We chose the defensive escort team problem with these environments, because it is a representative case of a well structured dual-reward collaborative MARL problem, the environments are continuous with a relatively high dimensional state space. Furthermore, previous work using these environments provide strong baselines against which the proposed algorithms can be compared. Aligning our experimental setup with~\cite{Sheikh-2019-COMPSAC, Sheikh-2019-AAMAS}, we chose \textit{4} bodyguards and \textit{10} bystanders.

\subsection*{Decomposing the Reward Function}
\label{sec:RewardFunctions}
In this section, we review the entangled multi-objective reward function defined in~\cite{Sheikh-2018-GoalsRL,Sheikh-2019-COMPSAC}, explain the problems with it and  decompose it to be used by DE-MADDPG.

As mentioned in the the previous section that the goal of the defensive escort team is to learn an optimal formation around the VIP to minimize the physical threat while simultaneously maintain a certain distance from the VIP to follow the social norms. To achieve both of these objectives, the multi-objective reward function for each agent $i$ is defined as:

\begin{equation}
\begin{aligned}
r_{total}  =& \alpha\overbrace{\left(\displaystyle -1+\prod_{k=1}^{M}\left(1-\mathit{RT}\left(VIP,b_{k}, R\right)\right) \right)}^{r_{global}} \\&
+(1-\alpha)\underbrace{\left( \mathcal{D}\left(\mathit{VIP},x_{i}\right)\right)}_{r_{local}}
\end{aligned}
\label{eq:composite}
\end{equation}

\noindent where $r_{global}$ represents the reward that each agent receives given the formation of the team around the VIP at time-step $t$, therefore, represents the main objective of the team and $r_{local}$ represents the reward that the agent receives for maintaining a certain distance from the VIP and is defined as 

\begin{equation}
\mathcal{D}\left(\mathit{VIP},x_{i}\right)=\begin{cases}
0 & m \leq\left\Vert x_{i}-\mathit{VIP}\right\Vert _{2}\leq d\\
-1 & \text{otherwise}\\
\\
\end{cases}
\end{equation}
\noindent where $m$ is the minimum distance the agent has to maintain from VIP and $d$ is the safe distance.
As per our understanding, $r_{total}$ has two problems. The first problem is the stability issue since the policy oscillates between optimizing the global reward and the local reward. This stability problem exacerbates when either of the rewards are sparse and the other reward is dense. This phenomenon is also explained in~\Cref{subsec:stability}. 

The second problem is the $\alpha$ hyper-parameter. The $\alpha$ hyper-parameter assigns weight to both rewards. Given the various difficulty settings of the simulations, security should be prioritized in some simulations as compared to the others. Moreover, as reinforcement learning experiments take substantial amount of time, finding an optimal $\alpha$ can be time consuming.  

We solve this problem by having a dual critic architecture where the task of the global centralized critic is to approximate the cumulative global reward while each agent's local critic approximates its own local reward. In this particular scenario, $Q^g_\psi$ learns to approximate $r_{global}$ while $Q^\pi_{\theta{_i}}$ approximates $r_{local}$. The benefit of this decomposition is more stable learning and exclusion of the $\alpha$ hyper-parameter.

\subsection{Evaluations}
\label{subsec:main}
To evaluate the efficacy of DE-MADDPG and its variants we compare our results with baseline MADDPG and DDPG on defensive escort team problem in the environments described above. We trained the global critic with two different approaches. In the first approach, we updated the policy by using the standard Multi-Agent Deep Deterministic Policy Gradient mentioned in~\Cref{eq:demaddpg} while in the second approach, we updated the policy using the Twin Delayed Deep Deterministic Policy Gradient mentioned in~\Cref{eq:demaddpg_td3}. Additionally, to mitigate the global sparse reward problem, we replace the standard replay buffer with prioritized experience replay buffer (PER). We performed our experiments on 8 different seeds. We used three layered neural networks  for both the critic and actor networks. For each environment, we trained each approach for $20,000$ episodes except the \textit{Pie-in-the-Face} environment which was trained for $10,000$ episodes.

~\Cref{fig:main_results} shows the learning curves of our experiments. ~\Cref{fig:random_landmark_reward} corresponds to the \textit{Random Landmark} environment and it can be seen that DE-MADDPG based approaches outperforms the MADDPG and DDPG by a significant margin. Similar observation can be seen for the other two environments i.e., \textit{Shopping Mall}, and \textit{Street}. Finally, for the last environment \textit{Pie-in-the-Face}, there is little difference between the performance of the different approaches. A possible reason for this is that this environment, focusing on a single attacking bystander, makes the positioning choice less complex.


Though, ~\cref{fig:main_results} shows the learning curves that visually represents the performance of the different approaches, it does not quantitatively explain the improvement in performance across different approaches. To that end, we test our trained policies across all environments on 8 different seeds. \Cref{table:results} shows the average returns of the global reward over $1000$ episodes. We notice that in complex environments such as \textit{Shopping Mall} and \textit{Random Landmarks}, DE-MADDPG augmented with TD3 and PER were able to achieve 55\% and 73\% better performance than baseline MADDPG. Similarly, on the slightly less complex environments \textit{Street} and \textit{Pie-in-the-Face}, the performance was about 61\% and 100\% better.

\ifbool{widetables}{
\begin{table*}[H]
\centering
\caption{Average return of the global reward over $1000$ episodes over 8 seeds. The best result each task is {\bf bolded}. $\pm$ corresponds to 95\% confidence interval over seeds.}
\label{table:results}
\begin{center}
\begin{small}
\begin{tabular}{lccccc}
\toprule
\bf{Environment} & \bf{DE-MADDPG(TD3+PER)} & \bf{DE-MADDPG(TD3)} & \bf{DE-MADDPG(PER)} & \bf{DE-MADDPG} & \bf{MADDPG}\\
\midrule
Shopping Mall 	& \bf{ -4.87 $\pm$ 0.06} & -5.61 $\pm$ 0.08		& -6.17 $\pm$ 0.08	 & -6.27 $\pm$ 0.07		& -7.59$\pm$ 0.13  \\
Random Landmarks & \bf{ -4.43 $\pm$ 0.05} & -4.91 $\pm$ 0.06 & -5.75 $\pm$ 0.04	 & -6.34 $\pm$ 0.08	& -7.67$\pm$ 0.13  \\
Street 		& \bf{-2.13 $\pm$ 0.06} & -2.38 $\pm$ 0.07 & -2.36 $\pm$ 0.07	 & -2.66 $\pm$ 0.08	& -3.43$\pm$ 0.13 \\
Pie-in-the-Face	& -0.07 $\pm$ 0.002 & \bf{-0.06 $\pm$ 0.002} & -0.11 $\pm$ 0.003	 & -0.07 $\pm$ 0.004	& -0.12$\pm$ 0.003 \\
\bottomrule
\end{tabular}
\end{small}
\end{center}
\end{table*}
}{
\begin{table}[H]
\caption{Average cumulative return of the global reward over 1000 episodes over 8 seeds. Maximum value for each task is bolded. $\pm$ corresponds to 95\% confidence interval over seeds.}
\label{table:results}
\adjustbox{width=\columnwidth}{
\begin{tabular}{p{2.1cm}p{2cm}p{2cm}p{2cm}p{2cm}p{1.8cm}p{1.8cm}}
\toprule
Environment & DE-MADDPG (TD3+PER) & DE-MADDPG (TD3) & DE-MADDPG (PER) & DE-MADDPG & MADDPG & DDPG\\
\midrule
Shopping Mall 	& \bf{ -4.87 $\pm$ 0.06} & -5.61 $\pm$ 0.08		& -6.17 $\pm$ 0.08	 & -6.27 $\pm$ 0.07		& -7.59$\pm$ 0.13  & -9.51$\pm$ 0.12\\
Rand. Landm. & \bf{ -4.43 $\pm$ 0.05} & -4.91 $\pm$ 0.06 & -5.75 $\pm$ 0.04	 & -6.34 $\pm$ 0.08	& -7.67$\pm$ 0.13  & -10.22$\pm$ 0.16\\
Street 		& \bf{-2.13 $\pm$ 0.06} & -2.38 $\pm$ 0.07 & -2.36 $\pm$ 0.07	 & -2.66 $\pm$ 0.08	& -3.43$\pm$ 0.13  	& -3.81$\pm$ 0.11\\
Pie-in-the-Face	& -0.07 $\pm$ 0.002 & \bf{-0.06 $\pm$ 0.002} & -0.11 $\pm$ 0.003	 & -0.07 $\pm$ 0.004	& -0.12$\pm$ 0.003 & -0.10$\pm$ 0.006\\
\bottomrule
\end{tabular}
} 
\end{table}
} 

\Cref{fig:main_results_local} and~\Cref{table:results_local} shows the learning curves and the test results of the sum of the local rewards. Similar to the global reward, it can be seen in~\Cref{fig:main_results_local} and and~\Cref{table:results_local} that DE-MADDPG based approaches outperforms MADDPG and DDPG results. One point to be noted here is that maintaining a certain distance from a moving payload is fairly an easy problem for standard reinforcement learning algorithms as the reward is dense and the state space is relatively easy but adding an additional objective such as global reward maximization not only had catastrophic affect on the global reward maximization but also negatively effected the learning of a trivial task. 

\begin{table}[h]
\caption{Average cumulative return of the local reward over 1000 episodes over 8 seeds. Maximum value for each task is bolded. $\pm$ corresponds to 95\% confidence interval over seeds.}
\label{table:results_local}
\adjustbox{width=\columnwidth}{
\begin{tabular}{p{2.1cm}p{2cm}p{2cm}p{2cm}p{2cm}p{1.8cm}p{1.8cm}}
\toprule
Environment & DE-MADDPG (TD3+PER) & DE-MADDPG (TD3) & DE-MADDPG (PER) & DE-MADDPG & MADDPG & DDPG\\
\midrule
Shopping Mall 	 &  -0.41 $\pm$ 0.02       & \bf{-0.23 $\pm$ 0.03}	 & -0.28 $\pm$ 0.03	     & -0.30 $\pm$ 0.07		& -1.44$\pm$ 0.07  	& -0.92$\pm$ 0.05 \\
Rand. Landm.     & \bf{ -0.15 $\pm$ 0.02}  & -0.19 $\pm$ 0.01        & -0.25 $\pm$ 0.02	     & -0.22 $\pm$ 0.03	    & -1.46$\pm$ 0.06  	& -0.94$\pm$ 0.05\\
Street 		     & \bf{-0.12 $\pm$ 0.01}   & -0.15 $\pm$ 0.02        & -0.16 $\pm$ 0.02	     & -0.18 $\pm$ 0.08	    & -1.23$\pm$ 0.08  	& -0.47$\pm$ 0.04\\
Pie-in-the-Face	 & \bf{-0.11 $\pm$ 0.01}   & -0.28 $\pm$ 0.01        & -0.12 $\pm$ 0.01	     & -0.31 $\pm$ 0.01	    & -0.20$\pm$ 0.01 	& -0.31$\pm$ 0.01 \\
\bottomrule
\end{tabular}
} 
\end{table}

A common pattern can be seen in~\Cref{fig:main_results} and~\Cref{fig:main_results_local} that DE-MADDPG based approaches not only achieve higher global and local rewards, they achieve it significantly faster as compared to  MADDPG and DDPG. For example, in complex environments such as \textit{Random Landmarks} and \textit{Shopping Mall}, DE-MADDPG (TD3 + PER) reaches the maximum performance before $2500$ episodes while the baseline MADDPG reaches its best performance for \textit{Random Landmarks} and \textit{Shopping Mall} environment at around $12,000$ episodes and $20,000$ episodes respectively.

\subsection{Stability}
\label{subsec:stability}
\begin{figure}
        \centering
        \includegraphics[height=2in, width=3.40in]{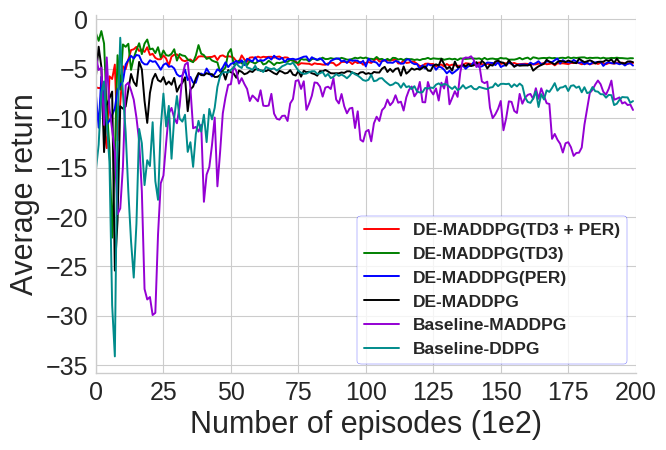}
\caption{Visualizing the single seed learning graph of shopping mall environment. Notice that DE-MADDPG(TD3) almost becomes flat after 10,000 episodes.}
\label{fig:stability}
\end{figure}

One of the focal point of this study was to find a stable solution that solves the dual-reward MARL problem. In this section, we analyse the stability of DE-MADDPG based solutions. Though,~\Cref{fig:main_results} shows that DE-MADDPG when augmented with TD3 outperforms MADDPG in maximizing global reward, the average of results across all the seeds make it difficult to analyze the stability. For that purpose, we chose fixed single seed runs of all the solutions on the \textit{Shopping Mall} environment and plotted their learning curves. It can be seen in~\Cref{fig:stability} that all DE-MADDPG based solutions are stable in learning to maximize global reward as compared to baseline MADDPG. This behavior is common across all seeds and environments except \textit{Pie-in-the-Face} as it is does not provide any complexity.

\begin{figure}[htb]
    \centering
    \begin{subfigure}[t]{0.25\textwidth}
        \centering
        \includegraphics[width=1.\linewidth, height=1.60in]{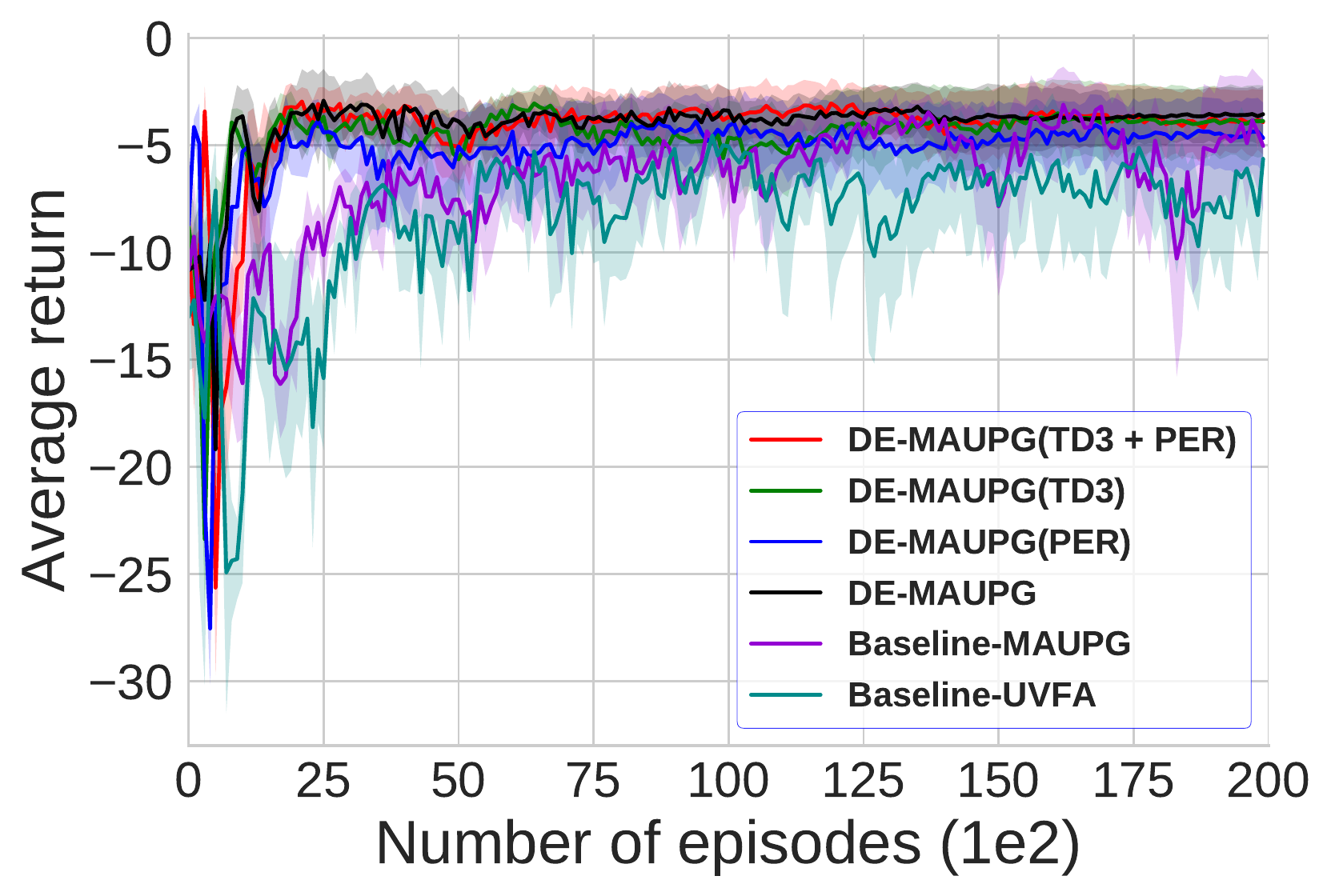}
        \caption{Random Landmarks}
        \label{fig:random_landmark_reward_maupg}
    \end{subfigure}%
    ~
    \begin{subfigure}[t]{0.25\textwidth}
        \centering
        \includegraphics[width=1.\linewidth, height=1.60in]{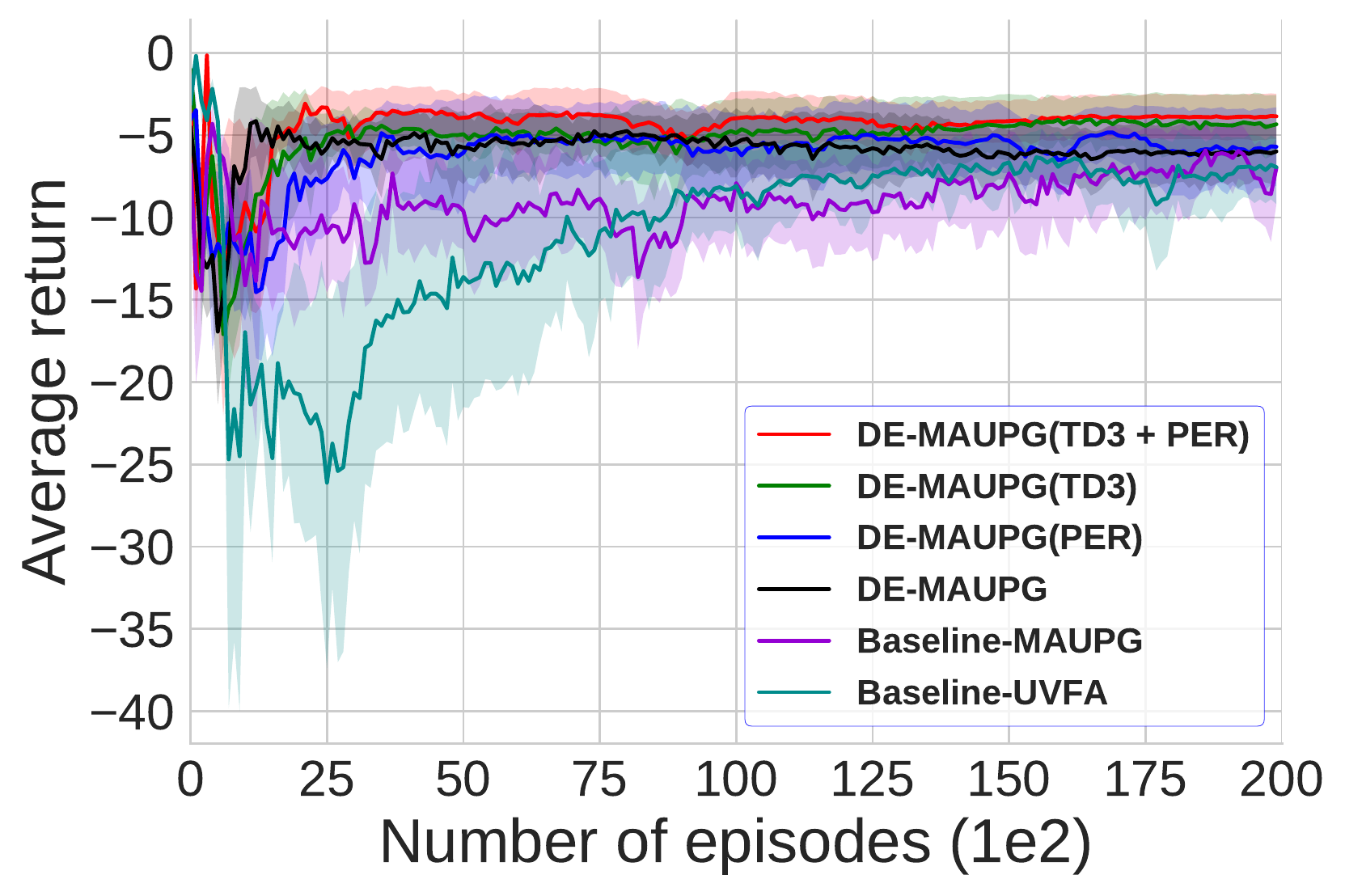}
        \caption{Shopping Mall}
        \label{fig:shopping_mall_reward_maupg}
    \end{subfigure}%
    \newline
    \begin{subfigure}[t]{0.25\textwidth}
        \centering
        \includegraphics[width=1.\linewidth, height=1.60in]{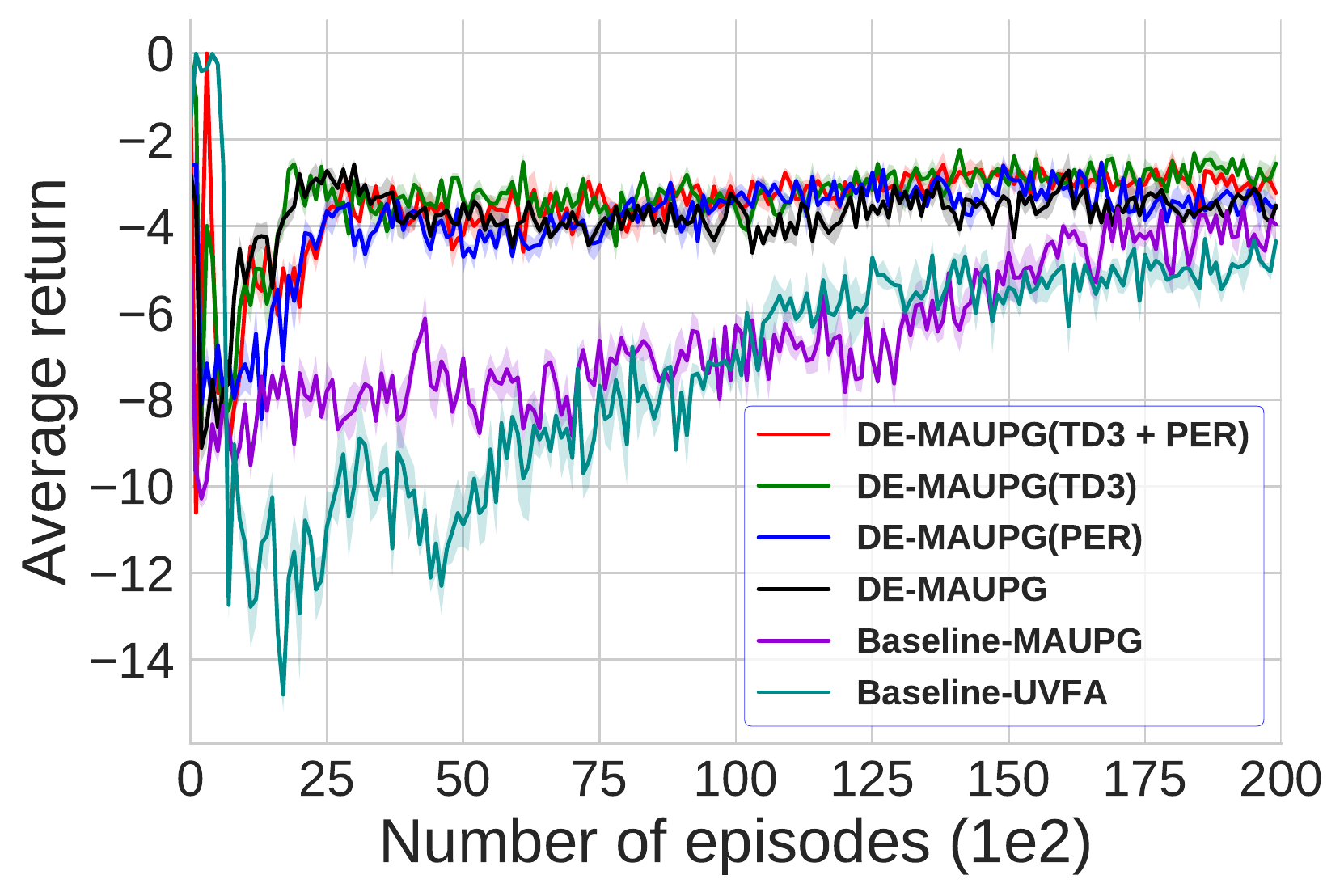}
        \caption{Street}
        \label{fig:street_reward_maupg}
    \end{subfigure}%
    ~
    \begin{subfigure}[t]{0.25\textwidth}
        \centering
        \includegraphics[width=1.\linewidth, height=1.60in]{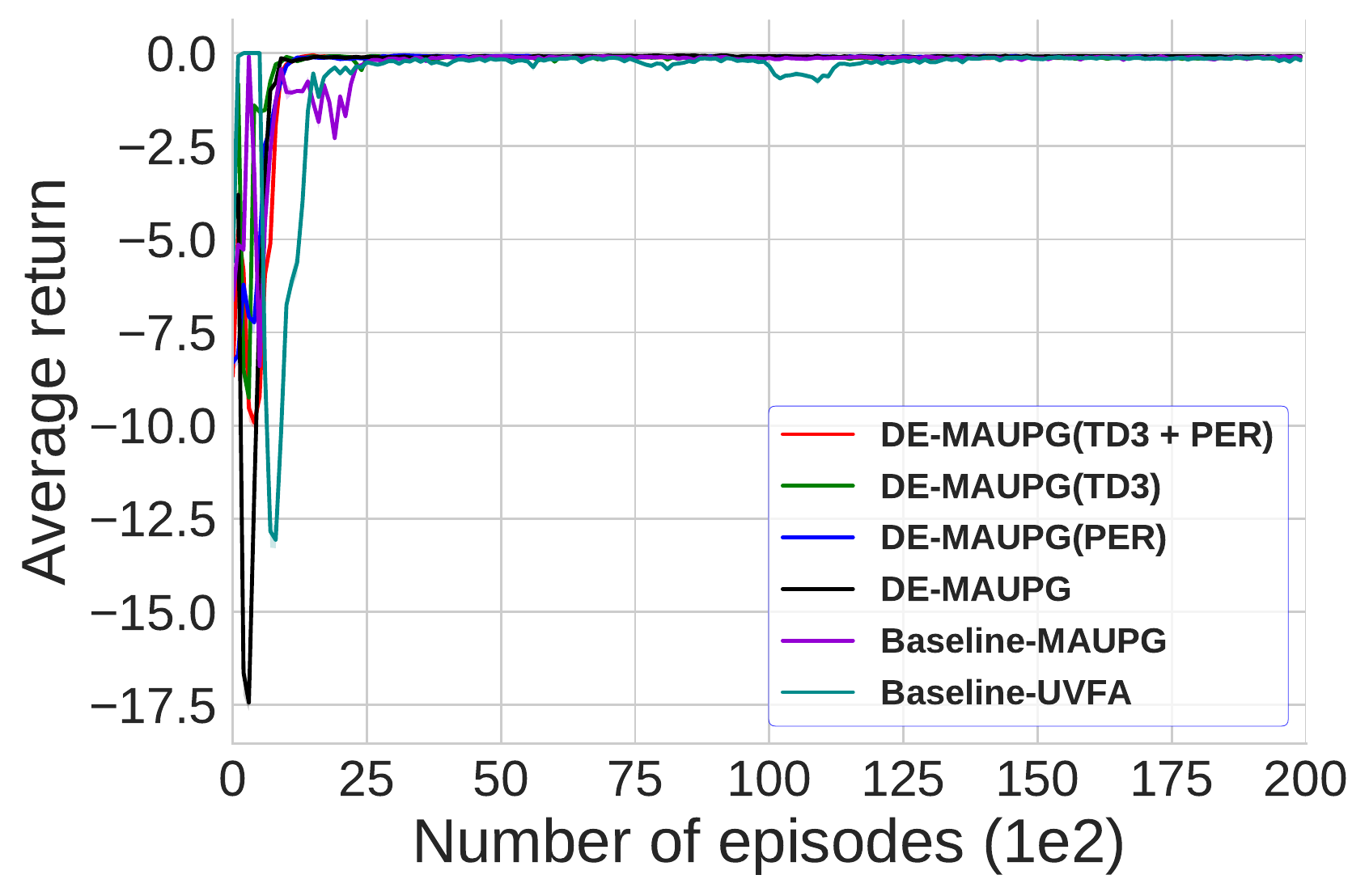}
        \caption{Pie-in-the-Face}
        \label{fig:red_carpet_reward_maupg}
    \end{subfigure}
    \caption{Learning curves representing the average cumulative  global reward of multi-scenario learning experiments..}
    \vspace{-5.5mm}
    \label{fig:main_results_maupg}
\end{figure}

\ifbool{widetables}{
\begin{table*}
\centering
\caption{Average Return of the global reward over 1000 episodes over 8 seeds. Maximum value for each task is bolded. $\pm$ corresponds to 95\% confidence interval over seeds.}
\label{table:results}
\begin{center}
\begin{small}
\begin{tabular}{lccccc}
\toprule
\bf{Environment} & \bf{DE-MAUPG(TD3+PER)} & \bf{DE-MAUPG(TD3)} & \bf{DE-MAUPG(PER)} & \bf{DE-MAUPG} & \bf{MAUPG}\\
\midrule
Shopping Mall 	& \bf{ -4.87 $\pm$ 0.06} & -5.61 $\pm$ 0.08		& -6.17 $\pm$ 0.08	 & -6.27 $\pm$ 0.07		& -7.59$\pm$ 0.13  \\
Random Landmarks & \bf{ -4.43 $\pm$ 0.05} & -4.91 $\pm$ 0.06 & -5.75 $\pm$ 0.04	 & -6.34 $\pm$ 0.08	& -7.67$\pm$ 0.13  \\
Street 		& \bf{-2.13 $\pm$ 0.06} & -2.38 $\pm$ 0.07 & -2.36 $\pm$ 0.07	 & -2.66 $\pm$ 0.08	& -3.43$\pm$ 0.13 \\
Pie-in-the-Face	& \bf{-0.07 $\pm$ 0.003} &  -0.13$\pm$ 0.008 & -0.08 $\pm$ 0.003	 & \bf{-0.07 $\pm$ 0.002}	& -0.12$\pm$ 0.003 \\
\bottomrule
\end{tabular}
\end{small}
\end{center}
\end{table*}
}{
\begin{table}
\caption{Average cumulative return of the global reward of multi-scenario learning over 1000 episodes over 8 seeds. Maximum value for each task is bolded. $\pm$ corresponds to 95\% confidence interval over seeds.}
\label{table:results_maupg}
\adjustbox{width=\columnwidth}{
\begin{tabular}{p{2.1cm}p{2cm}p{2cm}p{2cm}p{2cm}p{1.8cm}p{1.8cm}}
\toprule
Environment & DE-MAUPG (TD3+PER) & DE-MAUPG (TD3) & DE-MAUPG (PER) & DE-MAUPG & MAUPG & UVFA\\
\midrule
Shopping Mall 	& \bf{ -3.85 $\pm$ 0.07} & -4.34 $\pm$ 0.08		& -5.70 $\pm$ 0.09	 & -5.98 $\pm$ 0.10		& -6.97$\pm$ 0.11  & -6.94$\pm$ 0.11\\
Rand. Landm. &  -3.91 $\pm$ 0.07 & -3.86 $\pm$ 0.07 & -4.65 $\pm$ 0.08	 & \bf{-3.54 $\pm$ 0.08}	& -5.02$\pm$ 0.14  &-5.62$\pm$ 0.11 \\
Street 		& \bf{-2.54$\pm$ 0.08} & -3.22 $\pm$ 0.09 & -3.58 $\pm$ 0.11	 & -3.51 $\pm$ 0.10	& -3.97$\pm$ 0.14  &-4.34$\pm$ 0.14\\
Pie-in-the-Face	& \bf{-0.07 $\pm$ 0.003} &  -0.13$\pm$ 0.008 & -0.08 $\pm$ 0.003	 & \bf{-0.07 $\pm$ 0.002}	& -0.12$\pm$ 0.003 & -0.19$\pm$ 0.008\\
\bottomrule
\end{tabular}
} 
\end{table}
} 

\begin{table}[htb]
\caption{Average cumulative return of the local reward of multi-scenario learning over 1000 episodes over 8 seeds. Maximum value for each task is bolded. $\pm$ corresponds to 95\% confidence interval over seeds.}
\label{table:results_maupg_UVFA}
\adjustbox{width=\columnwidth}{
\begin{tabular}{p{2.1cm}p{2cm}p{2cm}p{2cm}p{2cm}p{1.8cm}p{1.8cm}}
\toprule
Environment & DE-MAUPG (TD3+PER) & DE-MAUPG (TD3) & DE-MAUPG (PER) & DE-MAUPG & MAUPG & UVFA\\
\midrule
Shopping Mall 	 &  -0.56 $\pm$ 0.02       & \bf{-0.27 $\pm$ 0.03}	 & -0.22 $\pm$ 0.03	     & -0.25 $\pm$ 0.07		& -1.04$\pm$ 0.06  	& -0.96$\pm$ 0.05 \\
Rand. Landm.     &  -0.18 $\pm$ 0.02  & -0.12 $\pm$ 0.02        & -0.22 $\pm$ 0.02	     & \bf{-0.11 $\pm$ 0.02}	    & -0.89$\pm$ 0.08  	& -0.71$\pm$ 0.08\\
Street 		     & -0.16 $\pm$ 0.02   & -0.15 $\pm$ 0.02        & -0.17 $\pm$ 0.02	     & \bf{-0.13 $\pm$ 0.02}	    & -0.64$\pm$ 0.07 	& -0.60$\pm$ 0.07 \\
Pie-in-the-Face	 & \bf{-0.21 $\pm$ 0.01}   & -0.43 $\pm$ 0.01        & -0.24 $\pm$ 0.01	     & -0.27 $\pm$ 0.01	    & -0.27$\pm$ 0.007 	& -0.22$\pm$ 0.009 \\

\bottomrule
\end{tabular}
} 
\end{table}

\subsection{Multi-Scenario Experiments}
Multi-agent reinforcement learning is sensitive to distortions and does not work well if the trained policies are deployed on scenarios other than the scenario on which the policies are trained on. This problem is generally referred as single-task multi-scenario learning. The goal here is to learn a joint policy $\boldsymbol{\pi}$ that performs equally well as scenario-dependant policy. In~\cite{Sheikh-2019-AAMAS} have introduced \textit{Multi-Agent Universal Policy Gradients} (MAUPG) to solve the multi-scenario learning and evaluated it on the VIP protection environments similar to our experiments. The main idea behind MAUPG is to replace the standard centralized Q-functions in MADDPG with Universal Value Function Approximators (UVFA). Given the similarity between MADDPG and MAUPG, we replaced all the critics in DE-MADDPG with UVFAs. For brevity, we will refer our multi-scenario solution as \textit{Decomposed-Multi-Agent Universal Policy Gradients} (DE-MAUPG)
For our experiments, we kept our settings identical to DE-MADDPG experiments and trained it for 20,000 episodes. The point to note here is that unlike scenario-dependant training, where every scenario was trained for 20,000 episodes, all scenarios are trained in parallel using one joint policy $\boldsymbol{\pi}$.

~\Cref{fig:main_results_maupg} shows the learning curves of the DE-MAUPG and its variants. Similar to our results from~\Cref{subsec:main}, decomposition based learning solutions outperform the baseline MAUPG.  The point to be noted here is that not our solutions learn to achieve higher reward but they also learn faster. This can be easily seen in~\Cref{fig:random_landmark_reward_maupg} and~\Cref{fig:shopping_mall_reward_maupg} where DE-MAUPG (TD3 + PER) reaches the maximum performance in less than 2500 episodes where baselines MAUPG reaches its peak performance at 12,500 episodes in \textit{Random Landmarks} environment and does not even reach its peak performance before 19,000 episodes for the \textit{Shopping Mall} environment.

\Cref{table:results_maupg} quantifies the improvement of DE-MAUPG in maximzing global reward when compared to MAUPG and UVFA. We find that the decomposition based approaches achieve 81\% on the \textit{Shopping Mall} environment and 41\% on the \textit{Random Landmarks} environment. Similarly \Cref{table:results_maupg_UVFA} quantifies the improvement of DE-MAUPG in maximzing local reward  when compared to MAUPG and UVFA. 
\begin{figure}
    \centering
    \begin{subfigure}[t]{0.25\textwidth}
        \centering
        \includegraphics[width=1.\linewidth, height=1.60in]{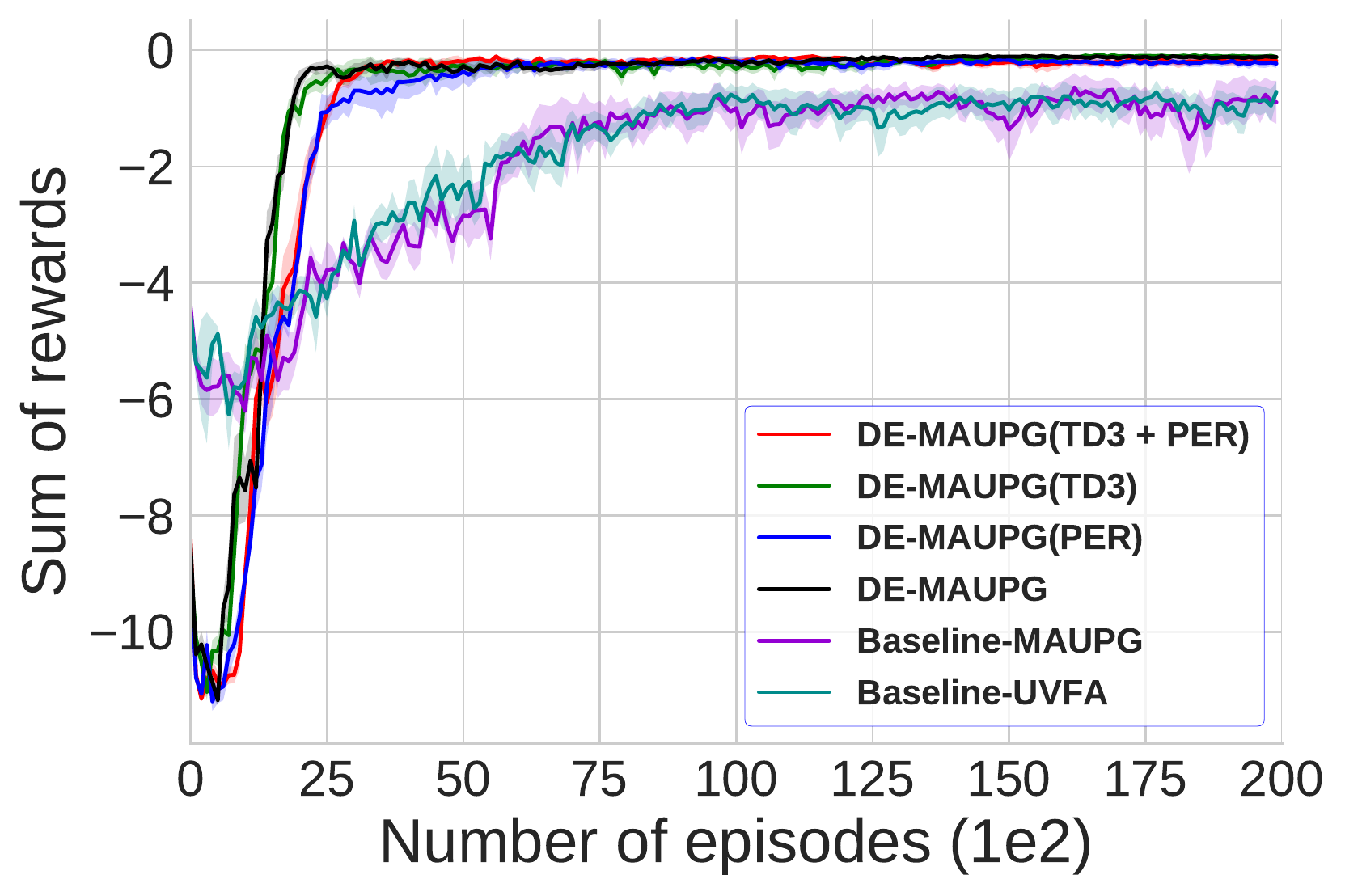}
        \caption{Random Landmarks}

        \label{fig:random_landmark_reward_maupg_local}
    \end{subfigure}%
    ~
    \begin{subfigure}[t]{0.25\textwidth}
        \centering
        \includegraphics[width=1.\linewidth, height=1.60in]{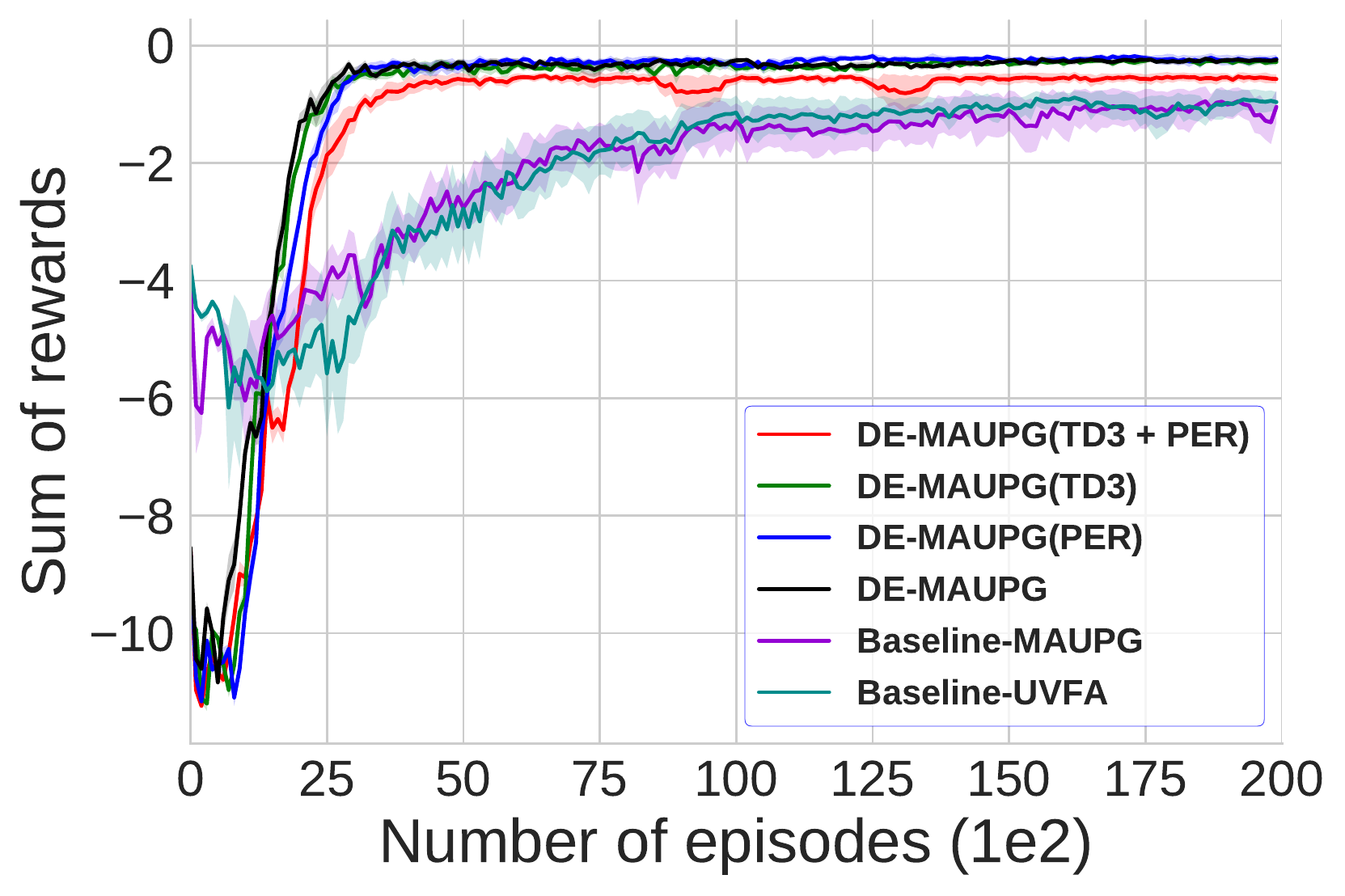}
        \caption{Shopping Mall}
        \label{fig:shopping_mall_reward_maupg_local}
    \end{subfigure}%
    \newline
    \begin{subfigure}[t]{0.25\textwidth}
        \centering
        \includegraphics[width=1.\linewidth, height=1.60in]{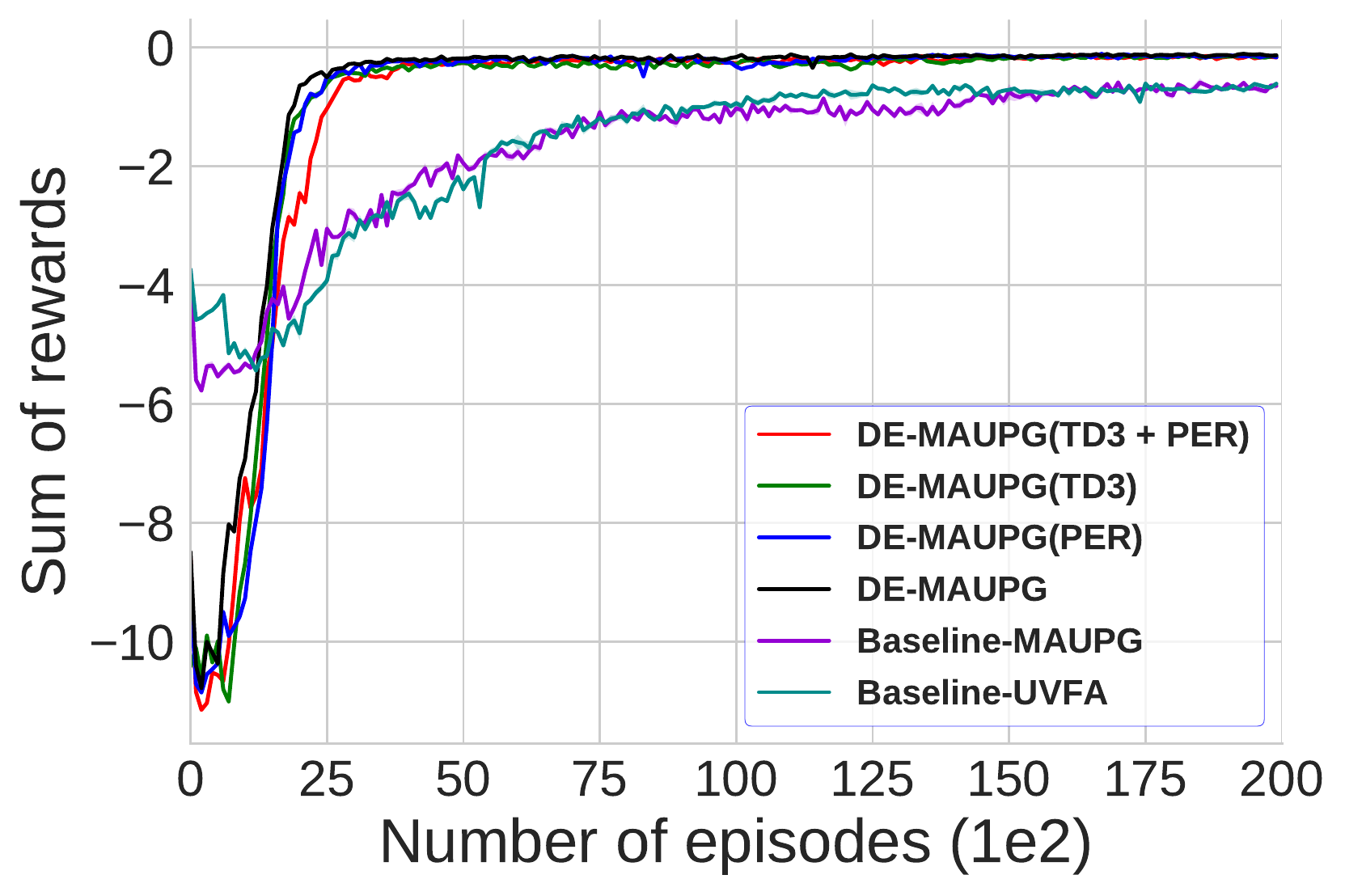}
        \caption{Street}
        \label{fig:street_reward_maupg_local}
    \end{subfigure}%
    ~
    \begin{subfigure}[t]{0.25\textwidth}
        \centering
        \includegraphics[width=1.\linewidth, height=1.60in]{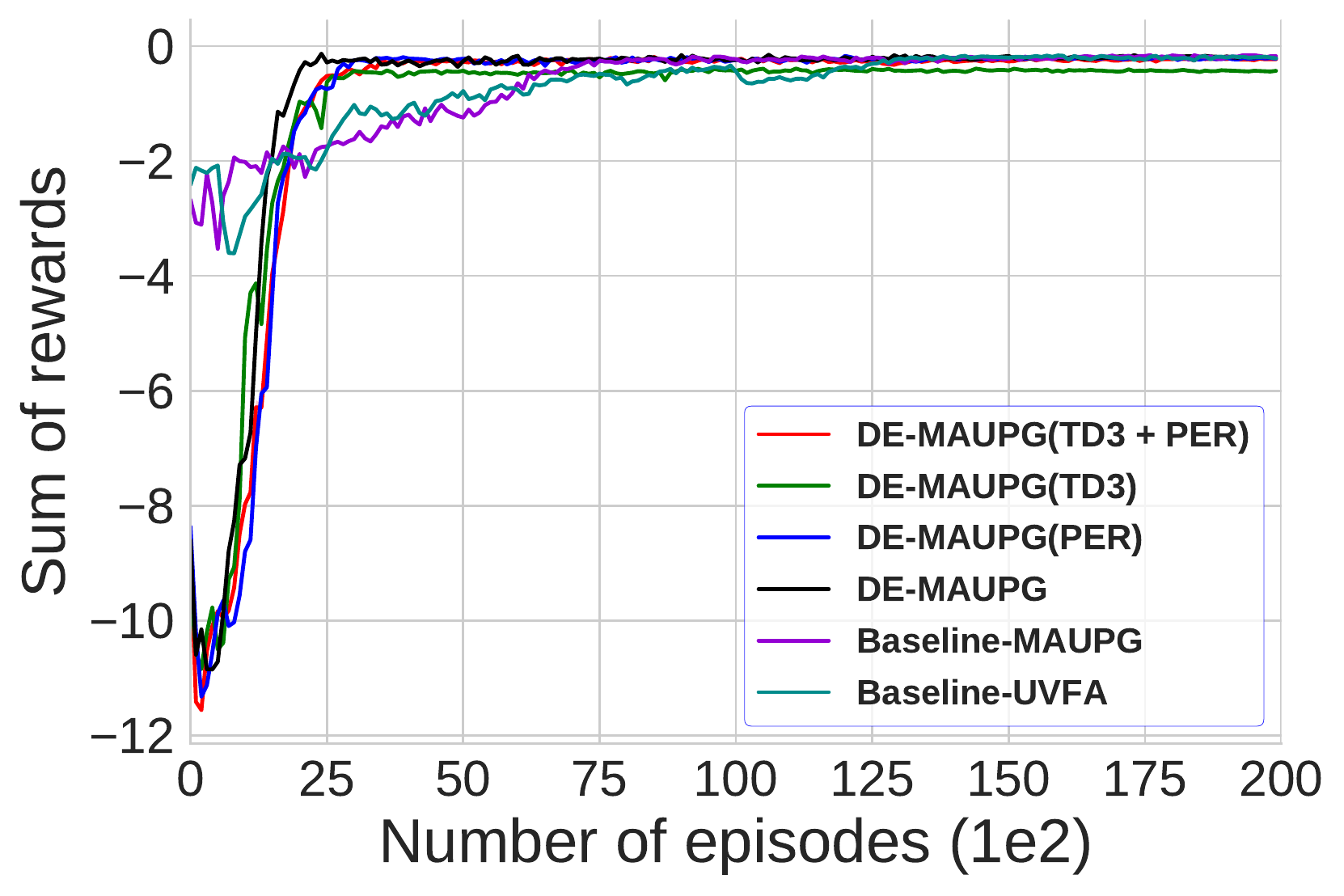}
        \caption{Pie-in-the-Face}
        \label{fig:red_carpet_reward_maupg_local}
    \end{subfigure}
    \caption{Learning curves representing the sum of local rewards of multi-scenario learning experiments.}
    \label{fig:main_results_maupg_local}
\end{figure}

\subsection{Computational Evaluations}
\label{subsec:compute}
\begin{figure}
        \centering
        \includegraphics[height=2in, width=3.40in]{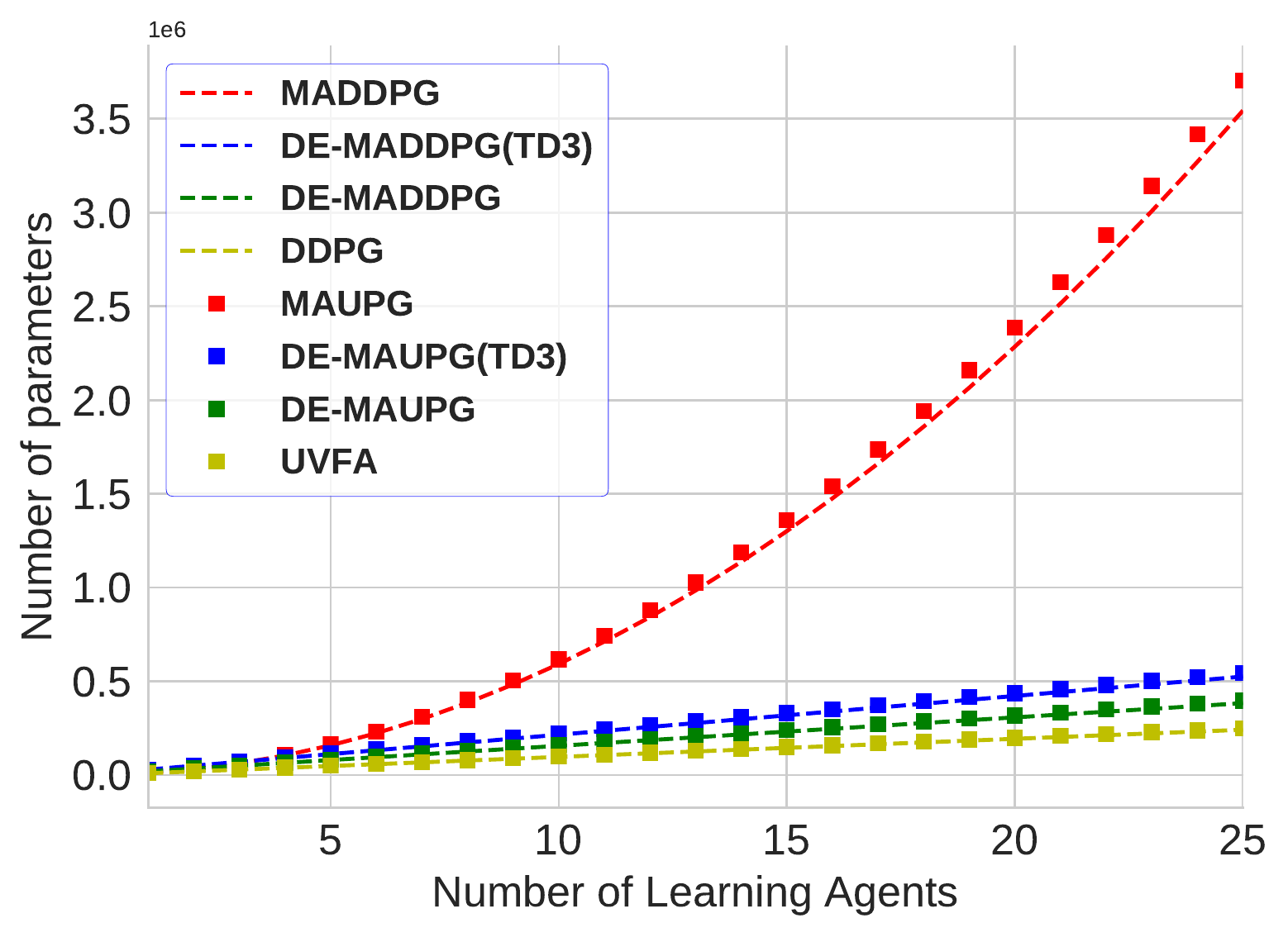}
\caption{Number of trainable parameters in the main Q-networks. Notice that 25 agents can be trained by DE-MADDPG approaches using the same number of parameters as compared to 10 agents if MADDPG is used.}
\label{fig:parameters}
\end{figure}

In this section we evaluate the growth of parametric space as the number of agents increases. Moreover, we empirically evaluate the computational time for DE-MADDPG and compare it with MADDPG and DDPG.
The main component of the MADDPG that fuels its learning are the distributed centralized Q-functions. As the number of the agents grow, the input space of those Q-functions increase quadratically. Concretely, assuming all the agents have identical observation and action space, the number of trainable parameters can be represented by $\mathcal{O}(n^2(odim+adim))$. Where $n$ is the number of agents, \textit{odim} and \textit{adim} represents the dimensionality of observation and action space respectively.  Alternatively, DE-MADDPG solves this scalability problem by having a shared global centralized Q-function whose parametric space increases linearly and can be represented as $\mathcal{O}(n(odim+adim))$. In~\Cref{fig:parameters}, we show the growth of number of parameters of all the main Q-networks of MADDPG, DE-MADDPG and its TD3 variant. It can be seen that our solution takes significantly small number of parameters as MADDPG and MAUPG. Note that the figure only represents the number of trainable parameters in the main Q-networks and does not include target or policy networks as the growth of the policy network parameters are identical and no gradient based learning happens in the target networks. All these statements hold true for their UVFA variants.

We empirically verified the benefits of parametric reduction by measuring the time taken to train the experiments. It can be seen in~\Cref{table:times} that DE-MADDPG based approaches always train faster than baseline MADDPG. The complete details about the network architecture can be seen in~\Cref{{sec:ExperimentDetails}}. We do not provide time comparisons for DDPG and UVFA variants as the experiments were run on multiple machines with a variety of hardware thus making it difficult for a fair comparison.

\ifbool{widetables}{
\begin{table*}[t]
\centering
\caption{Average training time for experiments in minutes over 8 different seeds. Minimum value for each task is bolded. $\pm$ corresponds to standard deviation across seeds}
\label{table:times}
\begin{center}
\begin{small}
\begin{tabular}{lccccc}
\toprule
\bf{Environment} & \bf{DE-MADDPG(TD3+PER)} & \bf{DE-MADDPG(TD3)} & \bf{DE-MADDPG(PER)} & \bf{DE-MADDPG} & \bf{MADDPG}\\
\midrule
Shopping Mall 	& 168.25 $\pm$ 7.22 &  152.85 $\pm$ 6.31		& 167.25 $\pm$ 4.40	 & \bf{149.5 $\pm$ 1.65}		& 268.5$\pm$ 21.21  \\
Random Landmarks & 168.87 $\pm$ 0.78 & 151.25 $\pm$ 1.19 & 164.14 $\pm$ 1.35	 & \bf{146.875 $\pm$ 1.16}	& 262.25$\pm$ 2.86  \\
Street 		& 397.62 $\pm$ 4.71 &  352.87 $\pm$ 5.63 & 356.87 $\pm$ 13.97	 & \bf{340.75 $\pm$ 16.74}	& 462.75$\pm$ 43.14 \\
Pie-in-the-Face	&  64.25  $\pm$ 1.23 &  59.25 $\pm$ 0.59 &  63.5$\pm$ 1.09	 & \bf{55.12 $\pm$ 1.05}	& 72.79$\pm$ 1.16 \\
\bottomrule
\end{tabular}
\end{small}
\end{center}
\vskip -0.1in
\end{table*}
}{
\begin{table}
\caption{Average training time for experiments in minutes over 8 different seeds. Minimum value for each task is bolded. $\pm$ corresponds to standard deviation across seeds}
\label{table:times}
\adjustbox{width=\columnwidth}{
\begin{tabular}{p{2cm}p{2cm}p{2cm}p{2cm}p{2cm}p{1.8cm}}
\toprule
Environ\-ment & DE-MADDPG\newline(TD3+PER) & DE-MADDPG\newline(TD3) & DE-MADDPG\newline(PER) & DE-MADDPG & MADDPG\\
\midrule
Shopping Mall 	& 168.25 $\pm$ 7.22 &  152.85 $\pm$ 6.31		& 167.25 $\pm$ 4.40	 & \bf{149.5 $\pm$ 1.65}		& 268.5$\pm$ 21.21  \\
Rand. Landm. & 168.87 $\pm$ 0.78 & 151.25 $\pm$ 1.19 & 164.14 $\pm$ 1.35	 & \bf{146.875 $\pm$ 1.16}	& 262.25$\pm$ 2.86  \\
Street 		& 397.62 $\pm$ 4.71 &  352.87 $\pm$ 5.63 & 356.87 $\pm$ 13.97	 & \bf{340.75 $\pm$ 16.74}	& 462.75$\pm$ 43.14 \\
Pie-in-the-Face	&  64.25  $\pm$ 1.23 &  59.25 $\pm$ 0.59 &  63.5$\pm$ 1.09	 & \bf{55.12 $\pm$ 1.05}	& 72.79$\pm$ 1.16 \\
\bottomrule
\end{tabular}
}
\end{table}
} 


\section{Experimental Details}
\label{sec:ExperimentDetails}

\ifbool{widetables}{
\begin{table*}[htb]
\caption{The parameters used for various variations of DE-MADDPG and the baseline algorithm MADDPG in the experiments.}
\label{table:parameters-1}
\adjustbox{width=\columnwidth}{
\begin{tabular}{p{2.1cm}p{2cm}p{2cm}p{2cm}p{2cm}p{1.8cm}p{1.8cm}}
\toprule
Parameter & DE-MADDPG (TD3+PER) & DE-MADDPG (TD3) & DE-MADDPG (PER) & DE-MADDPG & MADDPG & DDPG\\
\midrule
 Episodes & 20k & 20k & 20k & 20k & 20k & 20k \\
    Replay buffer & $10^6$ & $10^6$ & $10^6$ & $10^6$ & $10^6$  & $10^6$  \\
    Minibatch size & 2048 & 2048 & 2048 & 2048 & 2048 &2048\\
    Steps per train $Q_g$& 4 & 4 & 4 & 4 & N/A & N/A\\
    Steps per train $Q_l$& 2 & 2 & 2 & 2 & 2  & 2\\
    Max env steps & 25 & 25 & 25 & 25 & 25 &25\\
    PER $\alpha$  & 0.6 & N/A & 0.6 & N/A & N/A & N/A \\
    PER $\beta$  & 0.4 & N/A & 0.4 & N/A & N/A & N/A\\
    PER $\epsilon$  & 1e-6 & N/A & 1e-6 & N/A & N/A & N/A\\
    PER $\beta$ decay  & 10000 & N/A & 10000 & N/A & N/A & N/A\\

\bottomrule
\end{tabular}
} 
\end{table*}
}{ 
\begin{table}[htb]
\caption{The parameters used for various variations of DE-MADDPG and the baseline algorithm MADDPG in the experiments.}
\label{table:parameters-1}
\adjustbox{width=\columnwidth}{
\begin{tabular}{p{2.1cm}p{2cm}p{2cm}p{2cm}p{2cm}p{1.8cm}p{1.8cm}}
\toprule
Parameter & DE-MADDPG (TD3+PER) & DE-MADDPG (TD3) & DE-MADDPG (PER) & DE-MADDPG & MADDPG & DDPG\\
\midrule
  Episodes & 20k & 20k & 20k & 20k & 20k & 20k \\
    Replay buffer & $10^6$ & $10^6$ & $10^6$ & $10^6$ & $10^6$  & $10^6$  \\
    Minibatch size & 2048 & 2048 & 2048 & 2048 & 2048 &2048\\
    Steps per train $Q_g$& 4 & 4 & 4 & 4 & N/A & N/A\\
    Steps per train $Q_l$& 2 & 2 & 2 & 2 & 2  & 2\\
    Max env steps & 25 & 25 & 25 & 25 & 25 &25\\
    PER $\alpha$  & 0.6 & N/A & 0.6 & N/A & N/A & N/A \\
    PER $\beta$  & 0.4 & N/A & 0.4 & N/A & N/A & N/A\\
    PER $\epsilon$  & 1e-6 & N/A & 1e-6 & N/A & N/A & N/A\\
    PER $\beta$ decay  & 10000 & N/A & 10000 & N/A & N/A & N/A\\

\bottomrule
\end{tabular}
} 
\end{table}

} 

\subsection{Network Architecture}
Both actor and critic networks for all agents consists of 2 hidden layers containing 64 units in each layer. The hidden layers uses ReLU activation function while the output layers of both networks use linear activation function. Both networks are initialized using Xavier normal initializers  however,  the output layer of the target critics were initialized with uniform random values between (-0.01 and 0.01) to  enable one-step look ahead learning of the critics after each training cycle. To keep experiments as fair as possible, we initialized the network parameters with same seed across all the experiments.  Complete hyperparameter details can be seen in~\Cref{table:parameters-1}. Note that same configurations were used for multi-scenario learning experiments.
\section{Conclusions}
\label{sec:Conclusions}
In this paper, we focused on the dual-reward MARL: a collaborative setting where a group of learning agents have to simultaneously learn to maximize the collective global reward and individual local reward. To solve the problem,  we proposed the \textit{Decomposed Multi-Agent Deep Deterministic Policy Gradient (DE-MADDPG)} algorithm and applied it to the problem of defensive escort team: how can agent learn a policy to maintain an optimal formation around the VIP to protect him/her from possible physical attack. We first demonstrated that decomposing a multi-objective reward function leads to higher and more stable performance. We compared our results with the MADDPG algorithm and achieved at least 50\% better performance in terms of the collected reward. Additionally, we showed that our solution is computationally efficient and requires a significantly lower number of parameters while achieving better performance than the baseline. Finally, we showed that by replacing the standard critics with UVFAs, our solution also outperforms MAUPG which is a baseline algorithm for single-task multi-scenario learning in multi-agent reinforcement learning. 

\bibliographystyle{ieeetr}  
\bibliography{ref}  

\end{document}